# Evolutionary dynamics of complex traits in sexual populations in a strongly heterogeneous environment: how normal?

Léonard Dekens*†‡

11 December 2020

**Abstract**

When studying the dynamics of trait distribution of populations in a heterogeneous environment, classical models from quantitative genetics choose to look at its system of moments, specifically the first two ones. Additionally, in order to close the resulting system of equations, they often assume that the trait distribution is Gaussian (see for instance Ronce and Kirkpatrick 2001). The aim of this paper is to introduce a mathematical framework that follows the whole trait distribution (without prior assumption) to study evolutionary dynamics of sexually reproducing populations. Specifically, it focuses on complex traits, whose inheritance can be encoded by the infinitesimal model of segregation (Fisher 1919). We show that it allows us to derive a regime in which our model gives the same dynamics as when assuming a Gaussian trait distribution. To support that, we compare the stationary problems of the system of moments derived from our model with the one given in Ronce and Kirkpatrick 2001 and show that they are equivalent under this regime and do not need to be otherwise. Moreover, under this regime of equivalence, we show that a separation bewteen ecological and evolutionary time scales arises. A fast relaxation toward monomorphism allows us to reduce the complexity of the system of moments, using a slow-fast analysis. This reduction leads us to complete, still in this regime, the analytical description of the bistable asymmetrical equilibria numerically found in Ronce and Kirkpatrick 2001. More globally, we provide explicit modelling hypotheses that allow for such local adaptation patterns to occur.

## Introduction

Most species occupy heterogeneous environments, in which the spatial structure is expected to play a significant role in the evolution of the diversity of a species. As a result of the balance between the mixing effect of migration connecting the different habitats of a species and the selective pressure reducing diversity within each habitat, several equilibrium states encoding the local adaptation of a species can be reached. Will the species succeed to persist in a wide range of habitat available and thus thrive as a generalist species? Will it become adapted to specific sets of conditions as what we call a specialist species? Evolutionary biology fields have taken a sustained interest in these questions, in population genetics (Lythgoe 1997; Nagylaki and Lou 2001; Bürger and Akerman 2011; Akerman and Bürger 2014), adaptive dynamics (Meszéna, Czibula, and Geritz 1997; Day 2000) or quantitative genetics (Tufto 2000; Ronce and Kirkpatrick 2001; Hendry, Day, and Taylor 2001; Yeaman and Guillaume 2009; Débarre,

---

*Institut Camille Jordan, UMR5208 UCBL/CNRS, Université de Lyon, 69622 Villeurbanne, France.
†INRIA, Dracula Team
‡dekens@math.univ-lyon1.fr

Ronce, and Gandon 2013; Débarre, Yeaman, and Guillaume 2015; Mirrahimi 2017; Lavigne et al. 2019; Mirrahimi and Gandon 2020). Here we adopt the framework of quantitative genetics, which models the adaptation of a continuous trait without giving explicitly its underlying genetic architecture. Additionally, we specifically choose to analyse the influence of sexual reproduction as mating system.

**Model.** We build our model within a biological framework shared with classical studies (Ronce and Kirkpatrick 2001; Hendry, Day, and Taylor 2001; Débarre, Ronce, and Gandon 2013). We consider a sexual population whose individuals are characterized by a quantitative phenotypic trait $z \in \mathbb{R}$ and evolving in a heterogeneous environment constituted by two habitats that we will assume to be symmetric, illustrated in Fig. 1.

The density of population at a given time $t$ with respect to a phenotype $z$ in habitat $i \in \{1, 2\}$ is denoted $n_i(t, z) \in L^1(\mathbb{R}_+ \times \mathbb{R})$, for which we further assume that $z^k n_i(t, z) \in L^1(\mathbb{R}_+ \times \mathbb{R})$ for $k < 4$.

Local maladaptation is the source of mortality in our model: stabilizing selection acts quadratically in each patch toward an optimal phenotype $\theta_i \in \mathbb{R}$ with an intensity $g > 0$. Up to a translation in the phenotypic space, we can consider without loss of generality that $\theta_2 = -\theta_1 = \theta > 0$. Additionally, competition for resources regulates the total size of the subpopulation $N_i(t) = \int_{\mathbb{R}} n_i(t, z') \, dz'$ in each patch with an intensity $\kappa > 0$. The mortality rate of an individual with phenotypic trait $z \in \mathbb{R}$ is thus given by:

$$M[n_i(t, z)] = -g(z - \theta_i)^2 - \kappa N_i.$$

Migration between the two patches occurs symmetrically at a rate $m > 0$. The exchange of individuals from patch $i$ to patch $j$ of a given phenotype $z \in \mathbb{R}$ at time $t \geq 0$ is thereby:

$$m \left( n_j(t, z) - n_i(t, z) \right).$$

Finally, we denote by $\mathcal{B}_\sigma(n_i)(t, z)$ the number of new individuals that are born at time $t \geq 0$ in patch $i$ with a phenotype $z \in \mathbb{R}$ due to sexual reproduction. That phenomenon is occurring at a rate $r > 0$. The sexual reproduction operator is at this point still unspecified and will be defined below. However, we will consider that it respects the following conservative properties:

$$\forall t \in \mathbb{R}_+, \int_{\mathbb{R}} \mathcal{B}_\sigma(n_i)(t, z) \, dz = \int_{\mathbb{R}} n_i(t, z) \, dz, \quad \int_{\mathbb{R}} z \mathcal{B}_\sigma(n_i)(t, z) \, dz = \int_{\mathbb{R}} z \, n_i(t, z) \, dz.$$

The dynamics of the local trait distributions are therefore given by:

$$\begin{cases} \frac{\partial n_1}{\partial t}(t, z) = r\mathcal{B}_\sigma(n_1)(t, z) - g(z - \theta_1)^2 n_1(t, z) - \kappa N_1(t) n_1(t, z) + m \left( n_2(t, z) - n_1(t, z) \right), \\ \frac{\partial n_2}{\partial t}(t, z) = r\mathcal{B}_\sigma(n_2)(t, z) - g(z - \theta_2)^2 n_2(t, z) - \kappa N_2(t) n_2(t, z) + m \left( n_1(t, z) - n_2(t, z) \right). \end{cases} \quad (1)$$

**System of moments and gaussian assumption.** Quantitative genetics studies often model the dynamics of the sizes of the subpopulations $N_1 > 0$ and $N_2 > 0$ and their mean traits $\overline{z}_1$ and $\overline{z}_2$. Although we intend to follow the dynamics of the whole trait distributions, for the sake of comparison, we derive ordinary differential equations for the first moments of the trait distributions by integrating (1) with regard to $z$:

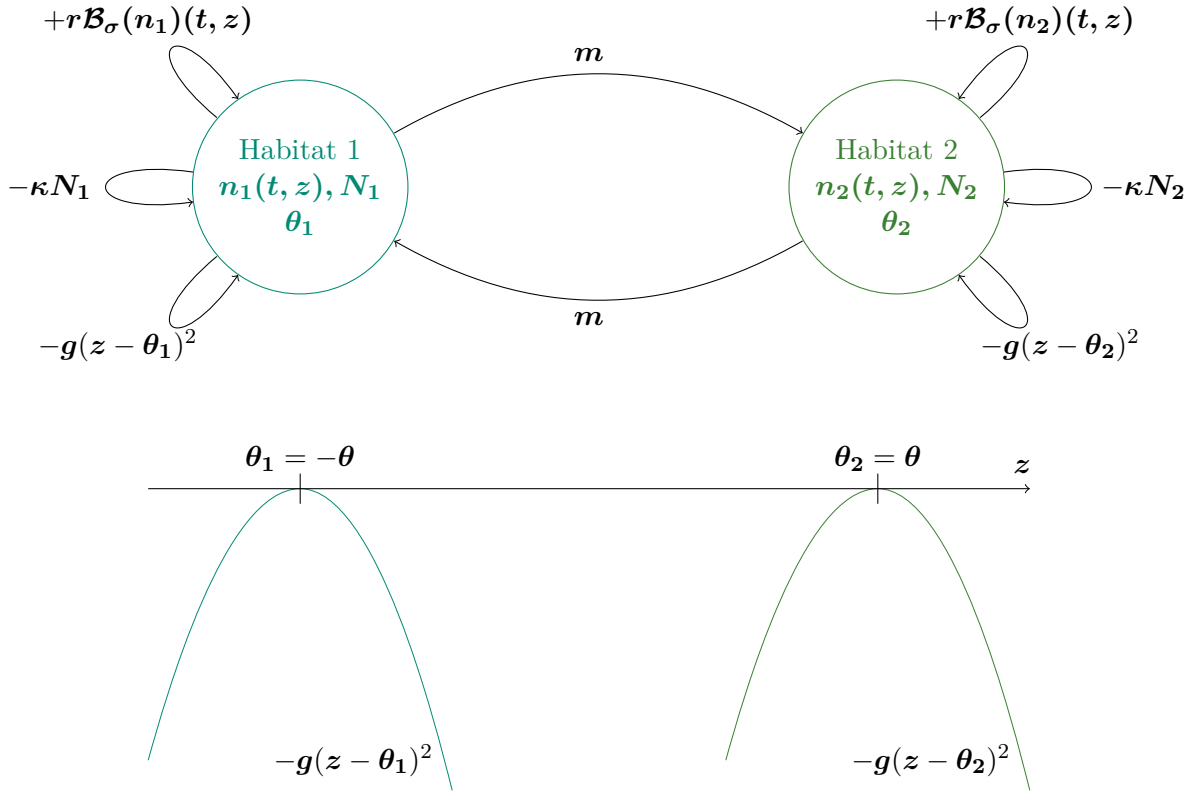

Figure 1: **Heterogeneous environment framework for a quantitative trait $z$.** The upper part of the figure illustrates the different biological forces acting in each habitat (reproduction, competition for resources, selection) and between them (migration). The lower part of the figure draws the local quadratic selection functions considered, where $\boldsymbol{\theta_1}$ and $\boldsymbol{\theta_1}$ are the local optimal traits.

$$\begin{cases} \frac{dN_i}{dt} = \left[r - \kappa N_i(t) - g(\overline{z}_i(t) - \theta_i)^2 - g\sigma_i^2\right] N_i(t) + m(N_j(t) - N_i(t)), \\ \frac{d\overline{z}_i}{dt} = 2\sigma_i^2 g(\theta_i - \overline{z}_i(t)) - g\psi_i^3 + m\frac{N_j(t)}{N_i(t)}(\overline{z}_j(t) - \overline{z}_i(t)). \end{cases} \quad (2)$$

where $\sigma_i^2$ and $\psi_i^3$ are respectively the variance and the third central moment of the trait distribution of each subpopulation (see Appendix A for details). At this point, a common key assumption used to close the system that arises in quantitative genetics models is the normality of such a trait distribution, with a constant variance (Hendry, Day, and Taylor 2001; Ronce and Kirkpatrick 2001). In Ronce and Kirkpatrick 2001, such an assumption results in the following system (with their original notations for the parameters):

$$\begin{cases} \frac{dN_i}{dt} = \left[r_0(1 - \frac{N_i}{K}) - \frac{\gamma}{2}\sigma_p^2 - \frac{\gamma}{2}(\overline{z_i} - \theta_i)^2\right] N_i + m(N_j - N_i), \\ \frac{d\overline{z_i}}{dt} = \sigma_g^2 \gamma(\theta_i - \overline{z_i}) + m\frac{N_j}{N_i}(\overline{z_j} - \overline{z_i}). \end{cases}$$

where $\sigma_p^2$ and $\sigma_g^2$ are respectively the constant phenotypic and genotypic variance, differing additively by a constant variance due to environmental effects $\sigma_e^2$ ($\sigma_p^2 = \sigma_g^2 + \sigma_e^2$). However, this approach disregards the effect of higher moments of the trait distribution (like the skewness), that may become significant due to the presence of gene flow, as pointed out in Yeaman and Guillaume 2009 and Débarre, Yeaman, and Guillaume 2015.

**The infinitesimal model of sexual reproduction.** To account for the influence of higher moments calls for models bypassing any prior assumption on the trait distribution, both to assess the validity of the Gaussian approximation or examine the departure from it. Therefore, it is necessary to make explicit the interplay between sexual reproduction and phenotypic inheritance. The infinitesimal model of sexual reproduction, first introduced by R.Fisher in 1919 (Fisher 1919) offers a simple way to tackle this issue for complex traits. Consequently, it has been used both in several biological studies (under truncation selection in Turelli and Barton 1994, or in a continent-island model in Tufto 2000) and mathematical ones (Mirrahimi and Raoul 2013; Bourgeron et al. 2017; Raoul 2017). Aligning with these, we choose it in our study to model trait inheritance due to sexual reproduction. The classical version of this model translates the stochasticity of the segregation process by the fact that the offspring trait variable $\mathcal{Z}$ (conditioned to the parental traits $\mathcal{Z}_1 = z_1$ and $\mathcal{Z}_2 = z_2$) follows a Gaussian law centered in the mean parental trait and with a segregational variance of $\frac{\sigma^2}{2}$:

$$\mathcal{Z}|\{\mathcal{Z}_1 = z_1, \mathcal{Z}_2 = z_2\} \sim \frac{z_1 + z_2}{2} + \mathcal{N}\left(0, \frac{\sigma^2}{2}\right). \quad (3)$$

Consequently, this model makes a normal assumption, not on the distribution of trait in the population, but on the distribution of offspring within each family, with a fixed and constant segregational variance (Turelli 2017). A common Mendelian interpretation of this mixing model is that the trait results from the expression of a large number of alleles with small additive effects (Fisher 1919; Bulmer 1971; Lange 1978). Recently, a rigorous framework of the use of that model in various biological contexts has been derived in Barton, Etheridge, and Véber 2017.

**The regime of small variance: $\sigma^2 \ll \theta^2$.** There also has been increasing mathematical interest in developing integro-differential equations for the whole trait distribution to study qualitatively quantitative genetics models (Magal and F. Webb 2000; Diekmann et al. 2005; Desvillettes et al. 2008). A framework introduced by Diekmann et al. 2005 to study asexual models in the regime of small mutation led to first rigorous results in Perthame and Barles 2008 in the context of homogeneous environment. Next, it has been extended to study spatially heterogeneous environment where asexual species evolve, like in Mirrahimi 2017 that successfully characterizes the equilibrium states by using a Hamilton-Jacobi approach in the limit of small mutations. For sexually reproducing populations, using the infinitesimal model in an asymptotic regime allowed Mirrahimi and Raoul 2013 to study invasions by phenotypically structured populations. More recently, using the infinitesimal model in a small variance regime led "Equilibria of quantitative genetics models beyond the Gaussian approximation I: Maladaptation to a changing environment" to formally derive features of the underlying trait distribution of a population under a changing environment. Their formal derivations have next been justified in a homogeneous space framework in Calvez, Garnier, and Patout 2019. Our work aligns with these studies: our main analysis lies in the small variance regime: $\sigma^2 \ll \theta^2$, namely when the diversity introduced by sexual reproduction is small compared to the heterogeneity of the environment.

**Contributions.** We use the infinitesimal model operator and the formalism of small segregational variance to study evolutionary dynamics of a sexually reproducing population under stabilizing selection in a heterogeneous and symmetrical environment in an integrated model (Section 1). From the PDE system on the local trait distributions, we derive a system of ODE on their moments. In the particular asymptotic regime considered: $\sigma^2 \ll \theta^2$, our ODE system approximates the one of Ronce and Kirkpatrick 2001 (Section 1):

$$\begin{cases} \frac{dN_i}{dt} = \left[r - \kappa N_i(t) - g(\overline{z}_i(t) - \theta_i)^2 - g\sigma^2\right] N_i(t) + m(N_j(t) - N_i(t)) + \mathcal{O}\left(\frac{\sigma^4}{\theta^4}\right), \\ \frac{d\overline{z}_i}{dt} = 2\sigma^2 g(\theta_i - \overline{z}_i(t)) + m\frac{N_j(t)}{N_i(t)}(\overline{z}_j(t) - \overline{z}_i(t)) + \mathcal{O}\left(\frac{\sigma^4}{\theta^4}\right). \end{cases} \quad (4)$$

To support that, we provide a numerical comparison between the two models, showing their equivalence in the small variance regime, and their discrepancy when this variance becomes large (Section 2). By doing so, we are justifying the validity of the Gaussian assumption on local trait distributions in this small variance regime. Next, we show that, in the regime of small variance, our system of moments can be reformulated as a slow-fast system (Section 3), which highlights the blending force of our sexual reproduction operator that strains monomorphism to quickly emerge at the meta-population level. The study of the corresponding unperturbed problem, with a reduced complexity, leads to the complete analytical description of the equilibria in the asymptotic regime of small variance. In particular, it gives the conditions of existence of bistable asymmetrical equilibria numerically observed by Ronce and Kirkpatrick 2001 (Section 4).

To replace this study in a broader context, let us first recall some findings of Ronce and Kirkpatrick 2001, our reference moment-based model in the quantitative genetics field. It makes a Gaussian assumption on the local trait distributions, without specifying any particular mode of reproduction. The authors numerically found that bistable mirrored asymmetrical equilibria can exist, allowing source-sink dynamics to completely reverse after a demographical loss event. Based on their study, however, it remains unclear which hypotheses on the inheritance process allow for such dynamics to arise. More recently, two studies interested in the equilibria states of asexual populations highlight the need for precise hypotheses with regard to such conclusions. If the authors of Débarre, Ronce, and Gandon 2013 indicate that asymmetrical equilibria can be locally stable in a restrained range of mutational parameters, Mirrahimi 2017 and Mirrahimi and Gandon 2020 show through using a continuum-of-alleles model that, under broader mutational parameters, only a single stable symmetrical equilibrium can arise in a symmetrical setting. Here, we claim that we can explain the dynamics of the analysis done in Ronce and Kirkpatrick 2001 via a model on phenotypic densities dynamics, analogous to Mirrahimi 2017 and Mirrahimi and Gandon 2020 but with a sexual reproduction operator derived from the infinitesimal model and in a small segregational variance regime. We thereby make explicit the details of another mechanism that can provide with those locally bistable asymmetrical equilibria, which relies on the blending effect of the infinitesimal model in a regime of small segregational variance.

# Contents

# 1  The infinitesimal model and the regime of small variance

In this section, we present the specific framework in which we choose to perform our analysis. We first present some properties of the infinitesimal model operator in general, then its relationship with the specific regime of small variance. Then, we will show that the asymptotic approximation allows us to formally derive a closed system for the dynamics of the moments.

Let us define the following rescaled variables and parameters to get a dimensionless system:

$$z := \frac{z}{\theta}, \quad g := \frac{g\theta^2}{r}, \quad m := \frac{m}{r}, \quad \varepsilon := \frac{\sigma}{\theta}, \quad t := \varepsilon^2 rt, \quad n_{\varepsilon,i}(t,z) := \frac{\kappa}{r} n_i(t,z),$$

and the reproduction operator $\mathcal{B}_\varepsilon(n_{\varepsilon,i})(t,z) = \mathcal{B}_\sigma(n_i)(t,z)$. Then, (1) gives the rescaled system:

$$\begin{cases} \varepsilon^2 \frac{\partial n_{\varepsilon,1}}{\partial t}(t,z) = \mathcal{B}_\varepsilon(n_{\varepsilon,1})(t,z) - g(z+1)^2 n_{\varepsilon,1}(t,z) - N_{\varepsilon,1}(t)n_{\varepsilon,1}(t,z) + m\left(n_{\varepsilon,2}(t,z) - n_{\varepsilon_1}(t,z)\right), \\ \\ \varepsilon^2 \frac{\partial n_{\varepsilon,2}}{\partial t}(t,z) = \mathcal{B}_\varepsilon(n_{\varepsilon,2})(t,z) - g(z-1)^2 n_{\varepsilon,2}(t,z) - N_{\varepsilon,2}(t)n_{\varepsilon,2}(t,z) + m\left(n_{\varepsilon,1}(t,z) - n_{\varepsilon,2}(t,z)\right). \end{cases} \quad (5)$$

From the remaining of this section and unless specified otherwise, we will refer to that system for all analysis purposes.

## 1.1  The sexual reproduction operator

**Presentation.**  For modelling the segregation process resulting from sexual reproduction, we use the infinitesimal model, first introduced in Fisher 1919. It is inspired originally from the observation that the phenotypic variance among families does not seem to depend on their breeding values (Galton 1877). Although this can be formulated solely from a phenotypic perspective, Fisher 1919 gives a Mendelian interpretation by proposing to consider that the quantitative trait $z$ results from the infinitesimally small additive effects of a large number of alleles. That interpretation, in the spirit of a central limit theorem, has been followed on (Bulmer 1971; Lange 1978; Bulmer 1980; Barton, Etheridge, and Véber 2017). It leads to (3). With our notations, we can express the number of individuals born at time $t$ with trait $z$ in habitat $i$ by:

$$\mathcal{B}_\varepsilon(n_{\varepsilon,i})(t,z) = \frac{1}{\sqrt{\pi}\varepsilon} \int_{\mathbb{R}^2} \exp\left[\frac{-(z - \frac{z_1+z_2}{2})^2}{\varepsilon^2}\right] n_{\varepsilon,i}(t,z_1) \frac{n_{\varepsilon,i}(t,z_2)}{N_{\varepsilon,i}(t)} dz_1 dz_2. \quad (6)$$

The scaled segregational variance $\frac{\varepsilon^2}{2}$ is assumed to be constant with regard to time and independent of the parental traits. These are strong biological assumptions. Their relevance in the context of a spatially structured population will be the subject of a forthcoming work.

**Equilibria under random mating only.** To study the behaviour of the reproduction operator (6), it is informative to consider the conservative case where a sexually reproducing population only experiences random mating, without any structure due to space or mating preferences:

$$\varepsilon^2 \frac{\partial n_\varepsilon}{\partial t}(t,z) = \frac{1}{\sqrt{\pi}\varepsilon} \int_{\mathbb{R}^2} \exp\left[\frac{-(z-\frac{z_1+z_2}{2})^2}{\varepsilon^2}\right] n_\varepsilon(t,z_1) \frac{n_\varepsilon(t,z_2)}{N_\varepsilon(t)} dz_1 dz_2 - n_\varepsilon(t,z), \qquad (7)$$

(the term $-n_\varepsilon(t,z)$ is meant to keep the size of the population constant by balancing birth and death). Then, every Gaussian distribution of variance $\varepsilon^2$ (arbitrarily centered) is a stable distribution under (7) (see Appendix B). Furthermore, it is shown in Raoul 2017 that there are no other equilibrium and that the convergence toward such a Gaussian distribution is exponential in quadratic Wasserstein distance. Therefore, with this operator of sexual reproduction, a fixed and finite variance in trait at equilibrium arises under random mating only and without selection.

## 1.2 The regime of small variance: $\varepsilon^2 \ll 1$.

The framework presented in this section is inspired by a methodology developed in Diekmann et al. 2005 and Perthame and Barles 2008 that uses asymptotic regime in partial differential equations in order to derive analytical features of quantitative genetics models. In a regime where few diversity is introduced by reproduction at each generation, the continuous trait distributions are expected to converge toward Dirac masses concentrated on some specific traits. Performing a suitable transformation on the trait distribution allows to unfold the singularities of these Dirac masses and define more regular objects to study and calculate, in order to follow trait densities. That methodology has already been successfully applied for asexual populations, in homogeneous (Perthame and Barles 2008) and heterogeneous space (Mirrahimi 2017), then in other frameworks such as the study of adaptation to a changing environment ("Equilibria of quantitative genetics models beyond the Gaussian approximation I: Maladaptation to a changing environment"), and lately for sexual populations in homogeneous space (Calvez, Garnier, and Patout 2019). Applying a similar approach as described above, we will show that, within a regime of small variance yet to be defined, we can reduce the complexity of the system while rigorously justify that reduction.

In our context, a relative measure of diversity introduced by reproduction comes from comparing the variance of the segregation process to a measure of habitats' difference:

$$\frac{\boldsymbol{\sigma}^2}{\boldsymbol{\theta}^2} = \varepsilon^2.$$

One can thus define the small variance regime by $\boldsymbol{\sigma}^2 \ll \boldsymbol{\theta}^2$, or equivalently $\varepsilon^2 \ll 1$. Moreover, we perform the unfolding of singularities by shaping the traits distributions according to:

$$n_{\varepsilon,i} = \frac{1}{\sqrt{2\pi}\varepsilon} e^{-\frac{U_{\varepsilon,i}}{\varepsilon^2}}. \qquad (8)$$

The exponential form, known as the Hopf-Cole transform in scalar conservation laws, presumes that $U_{\varepsilon,i}$ will be a more regular object to analyze when $\varepsilon^2 \ll 1$ than $n_{\varepsilon,i}$, which we expect to converge toward a sum of Dirac distributions centered at the minima of $U_{\varepsilon,i}$. In fact, "Equilibria of quantitative genetics models beyond the Gaussian approximation I: Maladaptation to a changing environment" performed a formal analysis on the behaviour of the reproduction term in the regime of small variance under such a formalism. They found that, for the various contributions to be well-balanced in the equation (reproduction and mortality) when $\varepsilon^2 \ll 1$, $U_{\varepsilon,i}$ is formally constrained to have the following expansion with regard to successive powers of $\varepsilon^2$ (see Appendix C):

$$U_{\varepsilon,i}(z) = \frac{(z - z_i^*)^2}{2} + \varepsilon^2 u_{\varepsilon,i}, \tag{9}$$

where $z_i^*$ is a byproduct of the formal analysis and $u_{\varepsilon,i}$ is the following order term in the expansion. It leads to:

$$n_{\varepsilon,i} = \frac{1}{\sqrt{2\pi}\varepsilon} e^{-\frac{(z-z_i^*)^2}{2\varepsilon^2}} e^{-u_{\varepsilon,i}(z)}. \tag{10}$$

Let us interpret this formalism. For $\varepsilon^2 \ll 1$, the leading term in the expansion (10) is precisely the Gaussian distribution of (yet unknown) mean $z_i^*$ and variance $\varepsilon^2$, namely a distribution we know to be at equilibrium under random mating only. Only considering this term would be to assume that the trait distribution is Gaussian. As we want to capture the departure from normality, we introduce the term $u_{\varepsilon,i}$, which we can see as the next order term in the expansion of $\log(n_{\varepsilon,i})$ with regard to successive powers of $\varepsilon$. It embodies the correction to the Gaussian distribution due to the effect of selection, competition and migration. The study of its analytical properties is beyond the scope of this paper and will be the project of a forthcoming paper. For now, we will assume that such a limit exist and we will use it in our analysis without rigorously justifying it.

### 1.3 Derivation of the dynamics of the moments in the regime of small variance

Although our method describes directly the trait distribution, we propose to formally derive the equations describing the dynamics of the first three moments of the trait distribution from its dynamics under the small variance of segregation ($\varepsilon^2 \ll 1$) to compare our framework to other quantitative genetic studies. Toward that purpose, we define (assuming persistence of each subpopulation):

$$\begin{aligned}
N_{\varepsilon,i}(t) &= \int_{\mathbb{R}} n_{\varepsilon,i}(t,z)\, dz, & \overline{z}_{\varepsilon,i}(t) &= \frac{1}{N_{\varepsilon,i}} \int_{\mathbb{R}} z\, n_{\varepsilon,i}(t,z)\, dz, \\
\sigma_{\varepsilon,i}^2(t) &= \frac{1}{N_{\varepsilon,i}} \int_{\mathbb{R}} (\overline{z}_{\varepsilon,i} - z)^2\, n_{\varepsilon,i}(t,z)\, dz, & \psi_\varepsilon^3 &= \frac{1}{N_\varepsilon} \int_{\mathbb{R}} (z - \overline{z}_{\varepsilon,i})^3 n_\varepsilon(z) dz.
\end{aligned} \tag{11}$$

Let us omit for a moment the time dependency. Using the expression (10) and under the formal assumption that $u := \lim_{\varepsilon \to 0} u_\varepsilon$ is sufficiently regular, we get the following expansions (where $v_{i,\varepsilon}$ is the expansion term of order $\varepsilon^4$ of $U_{\varepsilon,i}$ - see Appendix D):

$$\begin{cases}
N_{\varepsilon,i} = e^{-u_i(z_i^*)} \left[ 1 + \varepsilon^2 \left( \frac{(\partial_z u_i(z_i^*))^2}{2} - \frac{\partial_{zz} u_i(z_i^*)}{2} - v_{i,\varepsilon}(z_i^*) \right) \right] + \mathcal{O}(\varepsilon^4), \\
\overline{z}_{\varepsilon,i} = z_i^* - \varepsilon^2 \partial_z u_i(z_i^*) + \mathcal{O}(\varepsilon^4), \\
\sigma_{\varepsilon,i}^2 = \varepsilon^2 + \mathcal{O}(\varepsilon^4), \\
\psi_{\varepsilon,i}^3 = \mathcal{O}(\varepsilon^4).
\end{cases} \tag{12}$$

These expansions are informative, particularly the one describing the rescaled variance of the trait distribution. We can observe that it is equivalent to twice the rescaled segregational variance (which is given as a parameter of the model) when the latter is small. The local rescaled variance in trait are thereby asymptotically constant and independent of the local environment.

Now, from scaling (2) and using the formal expansions of the variances and skews given by (12) when $\varepsilon^2 \ll 1$ yields:

$$\begin{cases} \varepsilon^2 \frac{dN_{\varepsilon,i}}{dt} = \left[1 - N_{\varepsilon,i}(t) - g(\overline{z}_{\varepsilon,i}(t) - (-1)^i)^2 - g\varepsilon^2\right] N_{\varepsilon,i}(t) + m(N_{\varepsilon,j}(t) - N_{\varepsilon,i}(t)) + \mathcal{O}(\varepsilon^4), \\ \\ \varepsilon^2 \frac{d\overline{z}_{\varepsilon,i}}{dt} = 2\varepsilon^2 g((-1)^i - \overline{z}_{\varepsilon,i}(t)) + m\frac{N_{\varepsilon,j}(t)}{N_{\varepsilon,i}(t)}(\overline{z}_{\varepsilon,j}(t) - \overline{z}_{\varepsilon,i}(t)) + \mathcal{O}(\varepsilon^4), \end{cases} \quad (13)$$

which is equivalent to (4).

## 2 Equivalence with a moment based model

### 2.1 Presentation of the moment based model

In Ronce and Kirkpatrick 2001, the authors present a quantitative genetic model to tackle the same problem: the evolutionary dynamics of a species under the effects of stabilizing selection and migration between two symmetric patches. Let us first recall the model and indicate the parameters. Stabilizing selection toward a local phenotypic optima $\boldsymbol{\theta_i} \in \mathbb{R}$ is added to competition for resources within each patch to build the fitness of an individual of phenotype $\boldsymbol{z}$ in patch $i$:

$$\boldsymbol{r_i(z)} = \boldsymbol{r_0}\left(1 - \frac{\boldsymbol{N_i}}{\boldsymbol{K}}\right) - \frac{\boldsymbol{\gamma}}{2}(\boldsymbol{z} - \boldsymbol{\theta_i})^2,$$

where $\boldsymbol{r_0} > 0$ is the maximal fitness at low density, $\boldsymbol{K} > 0$ the carrying capacity of each environment (assumed to be the same in both of them), and $\boldsymbol{\gamma} > 0$ the intensity of the selection. Migration occurs symmetrically between the two patches at a rate $\boldsymbol{m} > 0$. The mode of reproduction is left unspecified, but phenotypes and breeding values are assumed to follow a Gaussian distribution within each population, of constant genetic ($\boldsymbol{\sigma_g}^2 > 0$) and phenotypic ($\boldsymbol{\sigma_p}^2 > 0$) variances, independent of the patch with:

$$\boldsymbol{\sigma_p}^2 = \boldsymbol{\sigma_g}^2 + \boldsymbol{\sigma_e}^2,$$

where $\boldsymbol{\sigma_e}^2 > 0$ is the environmental variance. The analysis is focused on the ordinary differential equation system of the first two moments of the local trait distributions (assuming persistence of each subpopulation). Namely, the sizes of the subpopulations ($\boldsymbol{N_1}, \boldsymbol{N_2}$) and the mean phenotypic traits ($\overline{\boldsymbol{z_1}}, \overline{\boldsymbol{z_2}}$):

$$\begin{cases} \frac{d\boldsymbol{N_i}}{dt} = \left[\boldsymbol{r_0}(1 - \frac{\boldsymbol{N_i}}{\boldsymbol{K}}) - \frac{\boldsymbol{\gamma}}{2}\boldsymbol{\sigma_p}^2 - \frac{\boldsymbol{\gamma}}{2}(\overline{\boldsymbol{z_i}} - \boldsymbol{\theta_i})^2\right]\boldsymbol{N_i} + \boldsymbol{m}(\boldsymbol{N_j} - \boldsymbol{N_i}), \\ \\ \frac{d\overline{\boldsymbol{z_i}}}{dt} = \boldsymbol{\sigma_g}^2 \boldsymbol{\gamma}(\boldsymbol{\theta_i} - \overline{\boldsymbol{z_i}}) + \boldsymbol{m}\frac{\boldsymbol{N_j}}{\boldsymbol{N_i}}(\overline{\boldsymbol{z_j}} - \overline{\boldsymbol{z_i}}). \end{cases} \quad (14)$$

### 2.2 Formal comparison

Let us consider (14) in the case where we neglect the additional variance due to the environment, so that all the variation in trait results from the genetic variance. We will denote this variance by $\varsigma^2$, so that: $\boldsymbol{\sigma_p}^2 = \boldsymbol{\sigma_g}^2 := \varsigma^2$. Then, let us also consider the equations of the trait distribution moments derived from our model (4), when disregarding the errors of $\mathcal{O}\left(\frac{\sigma^4}{\theta^4}\right)$. Then, the dynamics of the moments and their stationary states are equivalent under the change of parameters:

$$\boldsymbol{r} = \boldsymbol{r_0}, \quad \boldsymbol{g} = \frac{\boldsymbol{\gamma}}{2}, \quad \boldsymbol{\kappa} = \frac{\boldsymbol{r_0}}{\boldsymbol{K}}, \quad \sigma^2 = \varsigma^2, \quad \sigma_e = 0. \quad (15)$$

This change of parameters is only possible because, in both models, the variance in trait in the subpopulations is derived from a single parameter encoding the genetic stochasticity ($\sigma_g^2$ in Ronce and Kirkpatrick 2001 and $\sigma^2$ in our model). Particularly, the variance is independent from the other biological parameters, which is a structural difference with asexual models (see Mirrahimi 2017).

## 2.3 Numerical comparison

In this subsection, we provide results from numerical simulations performed to confirm this formal equivalence between the stationary states of the two models under the regime of small variance in which we expect this link to hold. In these simulations, we follow two systems:

- the first one is a discretization of (1), where we follow the evolution of the local trait distributions $n_i(t, \cdot)$. We then compute at each time the sizes, mean traits and variances in trait of the subpopulations $N_i(t)$, $\overline{z}_i(t)$ and $\sigma_i$. We emphasize the fact that we do not deduce $N_i(t)$ and $\overline{z}_i(t)$ from the system of moments (4).

- the second one is the system of moments (14) provided in the article Ronce and Kirkpatrick 2001, initialized by integration of $n_i(0, \cdot)$. We denote the respective quantitites $N_{i,RK}(t)$ and $\overline{z}_{i,RK}(t)$.

We then compare the evolution of the sizes and the mean traits of the subpopulations given by both systems. We also provide the evolution of the variance and the skewness in trait in both subpopulations compared to the value of the fixed and constant variance $\sigma_g$ and the skew null of the Gaussian approximation, for it can shed some lights on the divergence of the two systems. The results are displayed in Fig. 2.

**Parameters of the simulations.** The value of the parameters were taken from Ronce and Kirkpatrick 2001 (the optimal phenotypes are translated without loss of generality to reduce the numbers of parameters):

$$m = 0.1, \quad \gamma = 0.1, \quad r_0 = 1 - \frac{\gamma}{2}\sigma_p^2, \quad K = 2.42\, r_0, \quad \theta = |\frac{\theta_2 - \theta_1}{2}| = 3.5,$$

where the value of $\sigma_g^2 = \sigma_p^2 = \sigma^2$ determines completely the parameters. Two values are chosen for $\sigma^2 = \sigma_g^2 = \sigma_p^2$: the first, $\sigma^2 = 0.0025$, is set to assess the regime of small variance ($\sigma^2 \ll \theta^2$) in which our formal link of equivalence should hold. The second, $\sigma^2 = 1$, comes from the value set in Ronce and Kirkpatrick 2001 and illustrates the discrepancy between the two models when not in the small variance regime.

**Initial conditions.** In both simulations, the initial conditions are the same, conditioned to the value of $\sigma$, for we want to be close to the equilibrium when under random mating only and selection only, as if the two habitats were disconnected at first. We consider two populations locally adapted to their habitats, but one is a little smaller in size than the other. To do so, we set:

$$\begin{cases} n_1(0, z) = \frac{9}{10\kappa}\, \frac{e^{-\frac{(z+\theta)^2}{2\sigma^2}}}{\sqrt{2\pi}\sigma} \\ \\ n_2(0, z) = \frac{1}{\kappa}\, \frac{e^{-\frac{(z-\theta)^2}{2\sigma^2}}}{\sqrt{2\pi}\sigma} \end{cases}$$

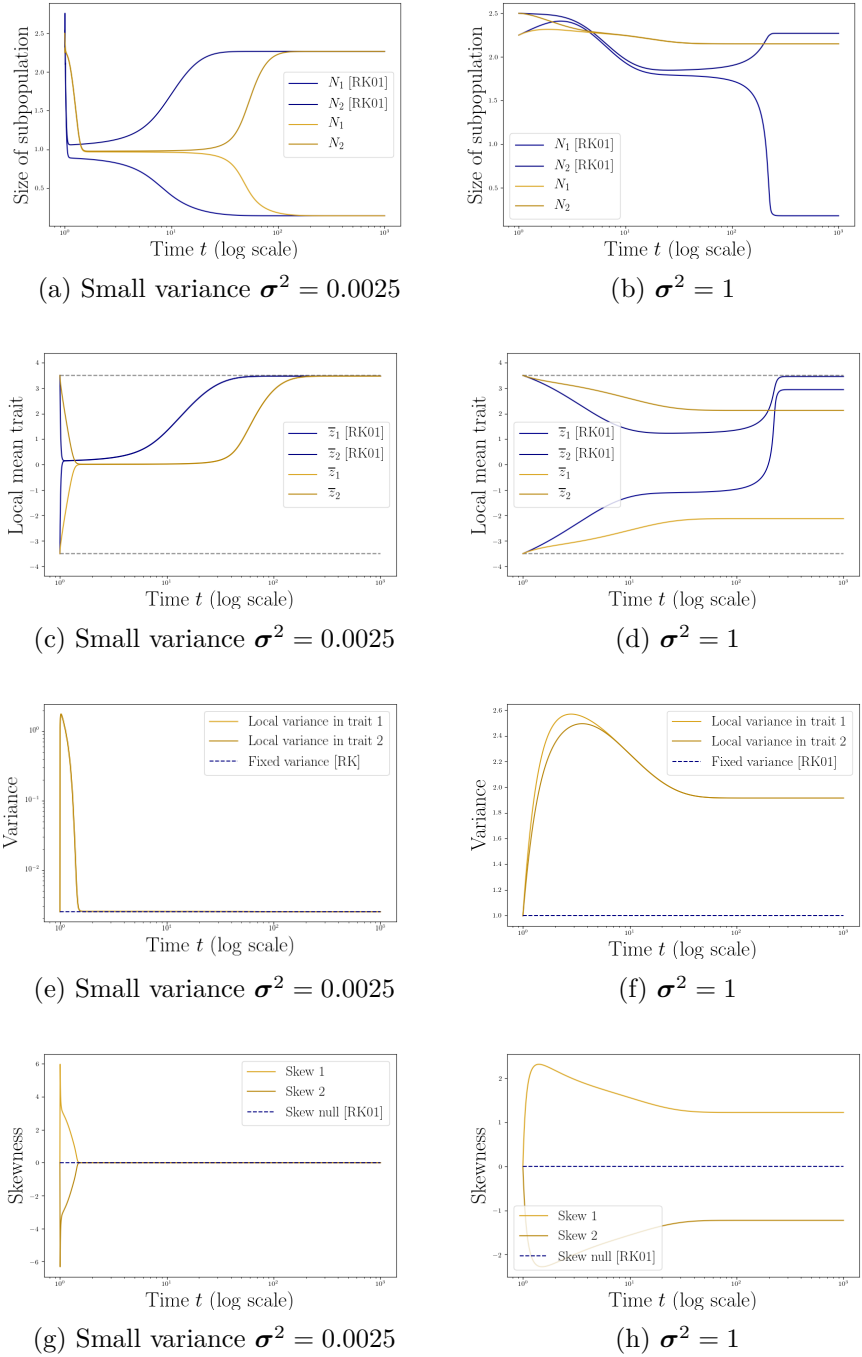

Figure 2: **Numerial comparison of our model (yellow lines) with Ronce and Kirkpatrick 2001's model (blue line) in small (left panel) and large (right panel) variance regime.** All parameters are the same or given by (15) and initial conditions are the same for both models. The left panel shows the results in the small variance regime ($\sigma^2 = 2.5 \times 10^{-3}$). Both models converge quickly to an asymmetrical equilibrium where both subpopulations are adapted to the second habitat ($\overline{z_1} = \overline{z_2} \approx \theta_2$). The right panel shows the results when not in the small variance regime ($\sigma^2 = 1$): the same link of equivalence does not hold. The discrepancy can be explained by looking at the local variances and skews in trait. They are asymptotically supporting a Gaussian assumption with fixed variance on the trait distributions in the small variance regime (note the logarithmic scale for the y-axis of Fig. 2e), but not when the segregational variance is larger. Note the logarithmic time scale for the sake of clarity.

**Results of the numerical comparison.** As Fig. 2a and Fig. 2c display the dynamics of the mean traits and population size in both subpopulations in the regime of small variance ($\sigma^2 = 0.0025$), it confirms numerically that both the model used in Ronce and Kirkpatrick 2001 and ours share similar dynamics (except maybe at initial times when the migratory fluxes are transiently high). When not in this regime ($\sigma^2 = 1$), Fig. 2b and Fig. 2d show that it does not need to be the case : the model used in Ronce and Kirkpatrick 2001 converges toward a monomorphic asymmetrical equilibrium whereas ours converges toward a dimorphic symmetrical equilibrium. The four bottom plots give an intuition of the source of this discrepancy. In the regime of small variance, we can see with Fig. 2e the variances in trait of the subpopulations in our model match the fixed genetical variance assumed by the gaussian approximation made in Ronce and Kirkpatrick 2001 (note the logarithmic scale for the y-axis on this figure). Moreover, Fig. 2g shows that the skew in both distributions are very small, as expected by our formal expansions, which makes the Gaussian approximation consistent. On the contrary, when not in the regime of small variance, Fig. 2f shows that the stationary variances in trait in both subpopulations derived from our model are significantly greater than the prescribed fixed variance $\sigma_g{}^2$ of Ronce and Kirkpatrick 2001. It is also important to note that with our model, even if the variance of segregation within families is held constant, the local variances in trait (byproducts of our numerical analysis) vary over time. The presence of respectively negative and positive skews (Fig. 2h) for the subpopulations confirms that the gaussian approximation breaks down in this regime in our model, hence the discrepancy in the outcomes with Ronce and Kirkpatrick 2001.

The two models have their own limit. Ronce and Kirkpatrick 2001 assumes that the variance in traits is the same in both subpopulations and constant through time and disregards any skewness in the local trait distributions. Our model assumption acts on the segregation : variance in each family is constant and independent of parental traits or habitat. As a result of that discrepancy between the models, their results differ on some ranges of parameters, as the previous figures show (Fig. 2b,Fig. 2d), while they match on others (Fig. 2a,Fig. 2c). To determine the range of parameters on which each model is closer to an explicit genetic model that includes drift, individual-based simulations are to be carried. That is the prospect of future work.

For now, since we have shown that our model was equivalent to Ronce and Kirkpatrick 2001's one in the regime of small variance, we will next develop a slow-fast analysis that will reduce the complexity of the system (Section 3) in the limit of vanishing variance in order to complete the equilbrium analysis done in Ronce and Kirkpatrick 2001 (Section 4).

## 3 Slow-fast system in small variance regime

In this section, we will see that the small variance regime allows for a slow-fast system to arise from (13). Using a singular perturbation approach similar to the one described in Levin and Levinson 1954, we will show that it converges in the limit of small variance to the following system, constrained in having $N_1^* > 0, N_2^* > 0$:

$$\begin{cases} \left[1 - N_1^* - g(z^* + 1)^2 - m\right] N_1^* + m N_2^* = 0, \\ \left[1 - N_2^* - g(z^* - 1)^2 - m\right] N_2^* + m N_1^* = 0, \\ \frac{dz^*}{dt} = 2g \left( \frac{\frac{N_2^*}{N_1^*} - \frac{N_1^*}{N_2^*}}{\frac{N_2^*}{N_1^*} + \frac{N_1^*}{N_2^*}} - z^* \right). \end{cases} \quad (16)$$

**Monomorphism in the regime of small variance.** This singular perturbation analysis reduces the complexity of the system (13) from four equations to three (see (16)), as the local mean traits $\bar{z}_{\varepsilon,1}$ and $\bar{z}_{\varepsilon,2}$ both relax rapidly toward the same value $z^*(t)$. Since asymptotically, the mean traits in both subpopulations are the same and the local variances in trait are infinitesimally small, the meta-population can be considered as <u>monomorphic</u> in $z^*(t)$, which we call the dominant trait. From the slow perspective (that can be seen as the evolutionary time scale), the two algebraic equations are defining a demographic equilibrium manifold on which this single dominant trait evolves until eventually reaching an evolutionary equilibrium (assuming persistence of population). Until further notice, let us consider ourselves in the regime of small variance: $\varepsilon^2 \ll 1$.

## 3.1 Slow-fast system formulation.

As we expect monomorphism to occur rapidly in the regime of small variance, let us operate the following change in variables:

$$\delta_\varepsilon = \frac{\bar{z}_{\varepsilon,2} - \bar{z}_{\varepsilon,1}}{2\varepsilon^2}, \quad z_\varepsilon^* = \frac{\bar{z}_{\varepsilon,2} + \bar{z}_{\varepsilon,1}}{2}.$$

Then (13) is equivalent to:

$$\begin{cases} \varepsilon^2 \frac{dN_{\varepsilon,1}}{dt} = \left[1 - N_{\varepsilon,1}(t) - g(z_\varepsilon^*(t) + 1 - \varepsilon^2\delta_\varepsilon(t))^2 - g\varepsilon^2\right] N_{\varepsilon,1}(t) + m(N_{\varepsilon,2}(t) - N_{\varepsilon,1}(t)) + \mathcal{O}(\varepsilon^4), \\[6pt] \varepsilon^2 \frac{dN_{\varepsilon,2}}{dt} = \left[1 - N_{\varepsilon,2}(t) - g(z_\varepsilon^*(t) - 1 + \varepsilon^2\delta_\varepsilon(t))^2 - g\varepsilon^2\right] N_{\varepsilon,2}(t) + m(N_{\varepsilon,1}(t) - N_{\varepsilon,2}(t)) + \mathcal{O}(\varepsilon^4), \\[6pt] \varepsilon^2 \frac{d\delta_\varepsilon(t)}{dt} = 2g - m\left(\frac{N_{\varepsilon,2}(t)}{N_{\varepsilon,1}(t)} + \frac{N_{\varepsilon,1}(t)}{N_{\varepsilon,2}(t)}\right)\delta_\varepsilon(t) + \mathcal{O}(\varepsilon^2), \\[6pt] \frac{dz_\varepsilon^*}{dt} = -2g z_\varepsilon^*(t) + m\left(\frac{N_{\varepsilon,2}(t)}{N_{\varepsilon,1}(t)} - \frac{N_{\varepsilon,1}(t)}{N_{\varepsilon,2}(t)}\right)\delta_\varepsilon(t) + \mathcal{O}(\varepsilon^2). \end{cases} \quad (17)$$

Let us denote $\Omega = (\mathbb{R}_+^*)^2 \times \mathbb{R}$ and $\bar{Y} = (N_1, N_2, \delta)$ the elements of $\Omega$. Let us define $F : \Omega \to \mathbb{R}$ and $G : \mathbb{R} \times \Omega \times [0,1] \to \mathbb{R}^3$ by :

$$\forall (z, (N_1, N_2, \delta), \varepsilon) \in \mathbb{R} \times \Omega \times [0,1],$$
$$F(N_1, N_2, \delta) = m\left(\frac{N_2}{N_1} - \frac{N_1}{N_2}\right)\delta,$$
$$G(z, N_1, N_2, \delta, \varepsilon^2) = \begin{pmatrix} \left[1 - N_1 - g(z + 1 - \varepsilon^2\delta)^2 - m - \varepsilon^2 g\right] N_1 + mN_2 \\ \left[1 - N_2 - g(z - 1 + \varepsilon^2\delta)^2 - m - \varepsilon^2 g\right] N_2 + mN_1 \\ 2g - \varepsilon^2 2g\delta - m\left(\frac{N_2}{N_1} + \frac{N_1}{N_2}\right)\delta \end{pmatrix}, \quad (18)$$

where $F$ and $G$ are respectively in $C^\infty(\Omega, \mathbb{R})$ and $C^\infty(\mathbb{R} \times \Omega \times [0,1], \mathbb{R}^3)$.

Let the following be called the perturbed system $(P_\varepsilon)$, where $\varepsilon > 0$ is a vanishing parameter and $\nu_{N,\varepsilon}$ and $\nu_{z,\varepsilon}$ are uniformly bounded as $\varepsilon \to 0$:

$$(P_\varepsilon) \begin{cases} \varepsilon^2 \frac{d\bar{Y}_\varepsilon}{dt} = G(z_\varepsilon, \bar{Y}_\varepsilon, \varepsilon^2) + \varepsilon^2 \nu_{N,\varepsilon}(t), \\ \frac{dz_\varepsilon}{dt} = -2g z_\varepsilon + F(\bar{Y}_\varepsilon) + \varepsilon^2 \nu_{z,\varepsilon}(t), \\ (z_\varepsilon(0), \bar{Y}_\varepsilon(0)) = (z_0^\varepsilon, \bar{Y}_0^\varepsilon). \end{cases} \quad (19)$$

One can verify that $(P_\varepsilon)$ is equivalent to (17). The framework is concordant with fast/slow system studies, like in Levin and Levinson 1954. We seek to establish the convergence over a finite time interval of the solutions of $(P_\varepsilon)$ towards the solution of the unperturbed system $(P_0)$, when $(z_0^\varepsilon, \bar{Y}_0^\varepsilon)$ is close enough to $(z_0^*, \bar{Y}_0^*)$ which verifies $G(z_0^*, \bar{Y}_0^*, 0) = 0$:

$$(P_0) \quad \begin{cases} G(z^*(t), \bar{Y}^*(t), 0) = 0, \\ \frac{dz^*}{dt} = -2gz^* + F(\bar{Y}^*) \\ (z^*(0), \bar{Y}^*(0)) = (z_0^*, \bar{Y}_0^*), \end{cases} \quad (20)$$

The first line $G(z^*(t), \bar{Y}^*(t), 0) = 0$ in (20) defines the slow manifold, parametrized by the slow variable $z^*(t)$, whereas the equation $\frac{dz^*}{dt} = -2gz^* + F(\bar{Y}^*)$ (second line) encodes the slow dynamic on that manifold. The slow manifold can be interpreted as the set of fast equilibria $\bar{Y}^*(t)$ corresponding to the levels given by slow variables $z^*(t)$. We will first assess the number of coexisting fast equilibria for any given parameter set $(g, m) \in \mathbb{R}_+^{*2}$ and value of the slow variable $z^*$. We will show that there exists either one or none of those, which constrains our proof of convergence to apply when $(z_0^\varepsilon, \bar{Y}_0^\varepsilon)$ is close enough to $(z_0^*, \bar{Y}_0^*)$ (the latter being on the slow manifold). Then, we will show that those fast equilibria are locally stable in Lemma 6. This lemma represents the essential condition for the convergence to apply on the finite time interval $[0, t^*]$, where $t^*$ will be subsequently defined (see Levin and Levinson 1954 and Appendix E for the detailed proof). We state the following theorem:

**Theorem 3.1.** *Let $(\bar{Y}^*, z^*)$ be solution of (20) on $[0, t^*]$ with initial conditions $(z_0^*, \bar{Y}_0^*)$, located on the slow manifold (ie. such that $G\left(z^*(t), \bar{Y}^*(t), 0\right) = 0$ for $t \in [0, t^*]$). For $0 < \varepsilon < 1$, let $(\bar{Y}_\varepsilon, z_\varepsilon)$ be solution of (19) on $[0, t^*]$ with initial conditions $(z_0^\varepsilon, \bar{Y}_0^\varepsilon)$. Then, as $\max(\varepsilon, |z_0^\varepsilon - z_0^*|, |\bar{Y}_0^\varepsilon - \bar{Y}_0^*|) \to 0$, $(\bar{Y}_\varepsilon, z_\varepsilon)$ converges toward $(\bar{Y}^*, z^*)$ uniformly on $[0, t^*]$.*

## 3.2 Number of coexisting fast equilibria.

Let us explicit that fast equilibria corresponding to $z^* \in \mathbb{R}$ are $\bar{Y}^* = (N_1^*, N_2^*, \delta^*) \in \Omega = (\mathbb{R}_+^*)^2 \times \mathbb{R}$ verifying: $G(z^*, \bar{Y}^*, 0) = 0$, ie. the system:

$$\begin{cases} [1 - N_1^* - g(z^* + 1)^2 - m] N_1^* + mN_2^* = 0, \\ [1 - N_2^* - g(z^* - 1)^2 - m] N_2^* + mN_1^* = 0, \\ 2g - m\left(\frac{N_2^*}{N_1^*} + \frac{N_1^*}{N_2^*}\right)\delta^* = 0. \end{cases} \quad (21)$$

We stress that this definition of fast equilibria requires both sizes of the subpopulations to be positive (we can notice that the two first equations of (21) do not allow for one population to go extinct while the other one persists). The objective is to identify how many coexisting fast equilibria there are for each set of parameter $(g, m, z^*) \in (\mathbb{R}_+^*)^2 \times \mathbb{R}$. To that purpose, let us first notice that the fast equilibria can be defined only using their demographic ratio $\frac{N_2^*}{N_1^*}$.

**Lemma 1.** *For $z^* \in \mathbb{R}$, let us define:*

$$P_{z^*}(X) = X^3 - f_1(z^*)X^2 + f_2(z^*)X - 1,$$

*where*

$$f_1(z^*) = 1 + \frac{g}{m}(z^* + 1)^2 - \frac{1}{m}, \quad f_2(z^*) = 1 + \frac{g}{m}(z^* - 1)^2 - \frac{1}{m}.$$

*If $(N_1^*, N_2^*, \delta^*)$ is a fast equilibrium, then: $\rho^* = \frac{N_2^*}{N_1^*}$ is a positive root of $P_{z^*}$ greater than $f_1(z^*)$. Conversely, if $\rho^*$ is a positive root of $P_{z^*}$ greater than $f_1(z^*)$, then:*

$$(N_1^*, N_2^*, \delta^*) = \left(m[\rho^* - f_1(z^*)], \, m\rho^*[\rho^* - f_1(z^*)], \, \frac{2g}{m\left(\rho^* + \frac{1}{\rho^*}\right)}\right) \in \Omega,$$

is a fast equilibrium corresponding to $z^*$ and $\rho^* = \frac{N_2^*}{N_1^*}$.

*Consequently, the number of fast equilibria corresponding to $z^*$ is the number of positive roots of $P_{z^*}(X)$ greater than $f_1(z^*)$.*

*Proof of Lemma 1.* For $z^* \in \mathbb{R}$, since $\bar{Y}^* \in \Omega = \mathbb{R}_+^* \times \mathbb{R}_+^* \times \mathbb{R}$, one can notice that (21) is equivalent to:

$$\begin{cases} \frac{N_2^*}{N_1^*} = \frac{g(z^*-1)^2 + m - 1 - m\frac{N_1^*}{N_2^*}}{g(z^*+1)^2 + m - 1 - m\frac{N_2^*}{N_1^*}}, \\ N_1^* = m\frac{N_2^*}{N_1^*} + 1 - g(z^*+1)^2 - m, \\ \delta^* = \frac{2g}{m\left(\frac{N_2^*}{N_1^*} + \frac{N_1^*}{N_2^*}\right)}. \end{cases}$$

$$\iff \begin{cases} \left[\frac{N_2^*}{N_1^*}\right]^3 - \left[\frac{N_2^*}{N_1^*}\right]^2 \left[1 + \frac{g}{m}(z^*+1)^2 - \frac{1}{m}\right] + \left[\frac{N_2^*}{N_1^*}\right]\left[1 + \frac{g}{m}(z^*-1)^2 - \frac{1}{m}\right] - 1 = 0, \\ N_1^* = m\left[\frac{N_2^*}{N_1^*} - \left(1 + \frac{g}{m}(z^*+1)^2 - \frac{1}{m}\right)\right], \\ \delta^* = \frac{2g}{m\left(\frac{N_2^*}{N_1^*} + \frac{N_1^*}{N_2^*}\right)}. \end{cases}$$

Hence the result. □

**Remark 3.1.** *Thanks to the symmetrical setting of the habitats, one can notice that, for all $z^* \in \mathbb{R}$, $P_{-z^*}(X) = X^3 P_{z^*}(1/X)$ and $f_1(-z^*) = f_2(z^*)$. Hence, the number of positive roots of $P_{z^*}$ that are greater than $f_1(z^*)$ is the number of positive roots of $P_{-z^*}$ that are greater than $f_2(z^*)$. Therefore, from now on, we will consider that $z^* \geq 0$ without loss of generality.*

The Lemma 2 shows that multiple fast equilibria cannot coexist and fast equilibria do not need to exist for any given set of parameters $(g, m, z^*) \in \mathbb{R}_+^{*2} \times \mathbb{R}_+$.

**Lemma 2.** *Let $z^* \geq 0$. Then:*

(i) *If $P_{z^*}$ has more than a single positive root, then they are all lower than $f_1(z^*)$. Hence, no fast equilibrium can exist in this configuration.*

(ii) *If $P_{z^*}$ has a single positive root $\rho^*$, then:*

$$[\rho^* > f_1(z^*)] \iff [f_1(z^*) \leq 0] \vee [P_{z^*}(f_1(z^*)) < 0].$$

*Proof of Lemma 2.* Let $z^* \geq 0$. As $P_{z^*}(0) = -1$, and the leading coefficient is 1, $P_{z^*}$ has at least one positive root and has either 1 or 3 positive roots.

(i) Let us assume that $P_{z^*}$ has three positive roots $x_1, x_2, x_3$. Then $f_1(z^*) = x_1 + x_2 + x_3 > \max\{x_1, x_2, x_3\}$, since the three roots are positive.

(ii) Let us assume now that $P_{z^*}$ has a single positive root $\rho^*$. As $P_{z^*}(0) = -1 < 0$ and the leading coefficient of $P_{z^*}$ is 1, we deduce that, for $y > 0$: $y < \rho^* \iff P_{z^*}(y) < 0$. Hence the result.

□

The second point of the Lemma 2 allows us to precise in the next proposition the conditions on $z^*$ such that a fast equilibrium exists, depending on $(g, m) \in \mathbb{R}_+^{*2}$ (see also Fig. 3):

**Proposition 3.1.** *For $(g, m, z^*) \in \mathbb{R}_+^* \times \mathbb{R}_+^* \times \mathbb{R}_+$ such that $P_{z^*}$ has a single positive root, let us define:*

$$\Delta = \frac{4}{g^2}\left[m^2 - 4g(m-1)\right], \quad z_1 = \frac{1}{2}\left[\frac{2(g+1-m)}{g} - \sqrt{\Delta}\right], \quad z_2 = \frac{1}{2}\left[\frac{2(g+1-m)}{g} + \sqrt{\Delta}\right].$$

*The following holds:*

* *If $g \geq 1$ and:*

  ◇ *$m < 2g\left(1 - \sqrt{1 - \frac{1}{g}}\right)$, then for all $z^* \in ]\sqrt{z_1}, \sqrt{z_2}[$, there exists a single fast equilibrium, and none otherwise.*

  ◇ *$m \geq 2g\left(1 - \sqrt{1 - \frac{1}{g}}\right)$ (ie. $\Delta \leq 0$), then for all $z^* \geq 0$, there exists no fast equilibria.*

* *If $g < 1$, then :*

  ◇ *If $m \leq \frac{1-g}{2}$, then, for $z^* \in [0, \sqrt{\frac{1-m}{g}} - 1[ \cup ]\sqrt{z_1}, \sqrt{z_2}[$, there exists a single fast equilibrium associated to $z^*$, and none otherwise.*

  ◇ *If $\frac{1-g}{2} < m < 1 - g$, then, for $z^* \in [0, \max\left(\sqrt{\frac{1-m}{g}} - 1, \sqrt{z_2}\right)[$, there exists a single fast equilibrium associated to $z^*$, and none otherwise.*

  ◇ *If $1-g \leq m$, then, for $0 \leq z^* < \sqrt{z_2}$, there exists a single fast equilibrium associated to $z^*$, and none otherwise.*

The proof of Proposition 3.1 is located in Appendix F.

Finally, we examine the conditions upon which $P_{z^*}$ has three positive roots. Due to the high degrees of the polynomials involved, an analytical condition on $(g, m) \in \mathbb{R}_+^{*\,2}$ has only been found when $z^* \in [-1, 1]$:

**Proposition 3.2.** *If $1 + 2m \geq g$, for all $z^* \in [-1, 1]$, $P_{z^*}$ has a single positive root.*

*If $1 + 2m < g$, there exists an interval $I \neq \emptyset$ centered in 0 such that for all $z^* \in I$, $P_{z^*}$ has three distinct positive roots.*

*Proof.* The proof will require three lemma. The first one states conditions upon which $P_{z^*}$ has three distinct positive roots for $z^* \in \mathbb{R}$. The second one gives an explicit condition determining if $P_0 = P_{z^*=0}$ has one ($1 + 2m \geq g$) or three distinct positive roots ($1 + 2m < g$). The third one shows that if there exists a $z^* \in [-1, 1]\setminus\{0\}$ such that $P_{z^*}$ has three distinct positive roots, then $P_0$ also has three distinct positive roots.

**Lemma 3.** *Let $z^* \in \mathbb{R}$. $P_{z^*}(X) = X^3 - f_1(z^*)X^2 + f_2(z^*)X - 1$ has three distinct positive roots if and only if the three following conditions hold simultaneously:*

(i) $f_1(z^*) > 0$,

(ii) $f_2(z^*) > 0$,

(iii) $\Delta(z^*) := f_1(z^*)^2 f_2(z^*)^2 - 4(f_1(z^*)^3 + f_2(z^*)^3) + 18f_1(z^*)f_2(z^*) - 27 > 0$.

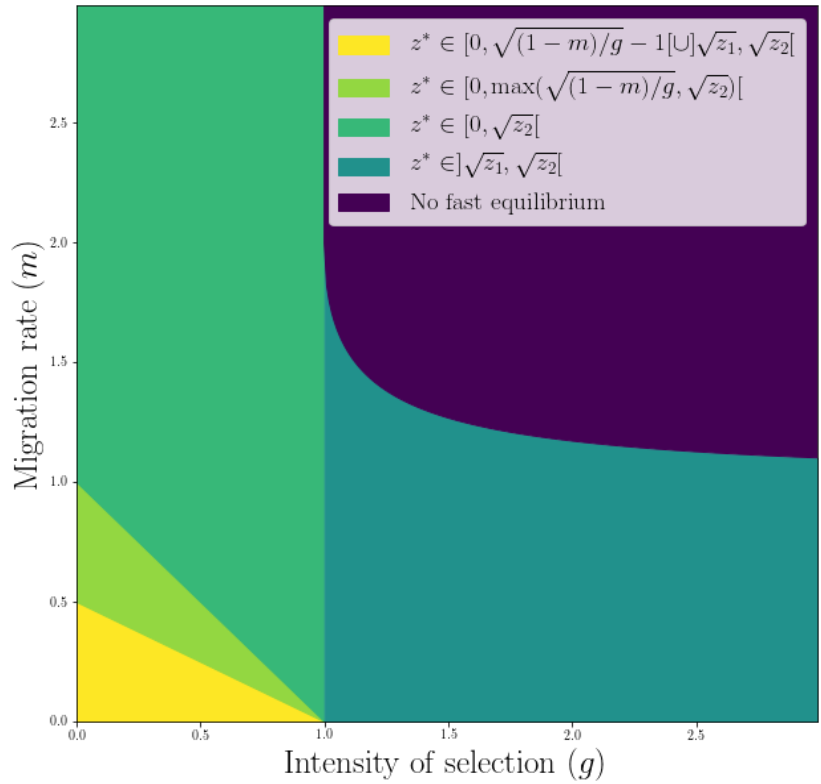

Figure 3: **Description of the conditions imposed on $z^* \geq 0$ to get a fast equilibrium depending on the pair $(g, m)$ under the preliminary assumption that $P_{z^*}$ has a single positive root, according to the results of Proposition 3.1.** When selection is smaller than 1, symmetrical fast equilibria exist ($z^* = 0$), and do not when selection is larger than 1. When both migration and selection are both too strong, no fast equilibrium can exist.

*Proof of Lemma 3.* Let $(x_1, x_2, x_3) \in \mathbb{C}^{*3}$ be the roots of $P_{z^*}$. Since $x_1 x_2 x_3 = 1$, we have:

$$f_1(z^*) = x_1 + x_2 + x_3, \quad f_2(z^*) = \frac{x_1 x_2 + x_2 x_3 + x_3 x_1}{x_1 x_2 x_3} = \frac{1}{x_1} + \frac{1}{x_2} + \frac{1}{x_3}.$$

Let us assume first that $x_1, x_2, x_3$ are positive and distinct. Then they are real and from the latter, $f_1(z^*) > 0$ and $f_2(z^*) > 0$. Moreover, they are real and distinct if and only if the discriminant of $P_{z^*}$ is positive, hence condition $(iii)$.

Conversely, let us assume $(i), (ii)$ and $(iii)$. Then $x_1, x_2, x_3$ are real and distinct. Since $P_{z^*}(0) < 0$, two of them (for example $x_2$ and $x_3$) share the same sign. Suppose that they are negative (they cannot be 0 since $P_{z^*}(0) = -1$) . Then $(i)$ yields:

$$x_1 > |x_2| + |x_3|.$$

Hence :

$$f_2(z^*) = \frac{1}{x_1} + \frac{1}{x_2} + \frac{1}{x_3} = \frac{1}{x_1} - \frac{1}{|x_2|} - \frac{1}{|x_3|} < \frac{1}{|x_2| + |x_3|} - \frac{1}{|x_2|} - \frac{1}{|x_3|} < 0,$$

which contradicts $(ii)$. Hence $x_1, x_2, x_3$ are positive and distinct. $\square$

**Lemma 4.** *$P_0 = P_{z^*=0}$ has three distinct positive roots if and only if $g > 1 + 2m$ and one positive root otherwise.*

*Proof of Lemma 4.* One can notice that $f_1(0) = f_2(0) = 1 + \frac{g}{m} - \frac{1}{m}$ and:

$$\Delta(0) = f_1(0)^4 - 8f_1(0)^3 + 18f_1(0)^2 - 27 = (f_1(0) + 1)(f_1(0) - 3)^3.$$

Hence, the precedent lemma ensures that $P_0$ has three distinct positives roots only in the region where $f_1(0) = f_2(0) = 1 + \frac{g}{m} - \frac{1}{m} > 0$ and $\Delta(0) > 0$. That occurs if and only if $f_1(0) > 3$, which yields $g > 1 + 2m$. $\square$

**Lemma 5.** *If there exists $z^* \in [-1, 1]$ such that $P_{z^*}$ has three distinct positive roots, then $P_0$ has three distinct positive roots.*

*Proof of Lemma 5.* We recall that we study the case $z^* > 0$ without loss of generality. Let us consider $z^* \in ]0, 1]$ such that $P_{z^*}$ has three distinct positive roots. From Remark 3.1, $P_{-z^*}$ has also three distinct positive roots. Thereby, Lemma 3 implies that $f_i(\pm z^*) > 0$, $i = 1, 2$ and $\delta(\pm z^*) > 0$.

It is clear that $f_1$ is strictly increasing on $]-1, 1[$ and $f_2$ is strictly decreasing on $]-1, 1[$. As $f_i(\pm z^*) > 0$, $i = 1, 2$, we get that $f_1 > 0$ and $f_2 > 0$ on $[-z^*, z^*]$, in particular $f_1(0) > 0$ and $f_2(0) > 0$.

Moreover, let us introduce the function $g : z \mapsto f_1(z)^2 - 3f_2(z)$. For $z \in ]-1, 1[$, $g'(z) = 2f_1'(z)f_1(z) - 3f_2'(z) > 0$, because $f_1(z) > 0$, $f_1'(z) > 0$ and $f_2'(z) < 0$. Therefore, $g$ is increasing on $]-1, 1[$. One can also notice that $g(z)$ is the quarter of the discriminant of $P_z'(X)$. As $P_{z^*}$ and $P_{-z^*}$ have three distinct positive roots, by Rolle's theorem, $P_{z^*}'$ and $P_{-z^*}'$ have two distinct positive roots. Therefore, $g(-z^*)$ and $g(z^*)$ are positive. As $g$ is increasing on $[-z^*, z^*]$, we get: $0 < g(0) = f_1(0)(f_1(0) - 3)$. Since $f_1(0) > 0$ and $g(0) > 0$, we have $3 < f_1(0) = 1 + \frac{g}{m} - \frac{1}{m}$. By the Lemma 4, $P_0$ has then three distinct positive roots.

$\square$

The successive applications of Lemma 4 and Lemma 5 are sufficient to conclude. $\square$

## 3.3 Fast relaxation towards the slow manifold.

We hereby prove the following lemma on the stability of the slow manifold:

**Lemma 6.** *For $(z, \bar{Y}) \in \mathbb{R} \times \Omega$ such that $G(z, \bar{Y}, 0) = 0$, $J_G(z, \bar{Y}) := \partial_{\bar{Y}} G(z, \bar{Y}, 0)$ is invertible. Furthermore, its eigenvalues are real and negative.*

*Proof.* For $(z, \bar{Y}) \in \mathbb{R} \times \Omega$ such that $G(z, \bar{Y}, 0) = 0$, we have:

$$J_G(z, \bar{Y}) = \begin{pmatrix} -2N_1 + [1 - g(z+1)^2 - m] & m & 0 \\ m & -2N_2 + [1 - g(z-1)^2 - m] & 0 \\ \frac{m\delta}{N_1}\left(\frac{N_2}{N_1} - \frac{N_1}{N_2}\right) & -\frac{m\delta}{N_2}\left(\frac{N_2}{N_1} - \frac{N_1}{N_2}\right) & -m\left(\frac{N_2}{N_1} + \frac{N_1}{N_2}\right) \end{pmatrix}.$$

Since $G(z, \bar{Y}, 0) = 0$, (18) leads to:

$$J_G(z, \bar{Y}) = \begin{pmatrix} -m\frac{N_2}{N_1} - N_1 & m & 0 \\ m & -m\frac{N_1}{N_2} - N_2 & 0 \\ \frac{m\delta}{N_1}\left(\frac{N_2}{N_1} - \frac{N_1}{N_2}\right) & -\frac{m\delta}{N_2}\left(\frac{N_2}{N_1} - \frac{N_1}{N_2}\right) & -m\left(\frac{N_2}{N_1} + \frac{N_1}{N_2}\right) \end{pmatrix}$$

$$= \begin{pmatrix} J & \begin{matrix} 0 \\ 0 \end{matrix} \\ \frac{m\delta}{N_1}\left(\frac{N_2}{N_1} - \frac{N_1}{N_2}\right) \quad -\frac{m\delta}{N_2}\left(\frac{N_2}{N_1} - \frac{N_1}{N_2}\right) & -m\left(\frac{N_2}{N_1} + \frac{N_1}{N_2}\right) \end{pmatrix}$$

so that we can compute:

$$\det J_G(z, \bar{Y}) = -m\left(\frac{N_2}{N_1} + \frac{N_1}{N_2}\right)\left[m\frac{N_2^2}{N_1} + m\frac{N_1^2}{N_2} + N_1 N_2\right] < 0.$$

Hence $J_G(z, \bar{Y})$ is invertible. A first eigenvalue is $-m\left(\frac{N_2}{N_1} + \frac{N_2}{N_1}\right) < -2m$. The last two eigenvalues are those of the upper left block $J$. We have:

$$\operatorname{tr}(J) < -2m < 0, \quad \det(J) = m\frac{N_1^2}{N_2} + m\frac{N_1^2}{N_2} + N_1 N_2 > 0,$$

and:

$$\operatorname{tr}(J)^2 - 4\det(J) = 4m^2 + \left(m\frac{N_2}{N_1} - m\frac{N_1}{N_2} + N_1 - N_2\right)^2 > 4m^2 > 0.$$

Hence $J$ has two real negative eigenvalues and consequently, $J_G(z, \bar{Y})$ has three real negative eigenvalues. □

# 4 Analytical description of the equilibria in the limit of vanishing variance

In this section, we will perform an equilibrium analysis for the stationary problem in the limit of vanishing variance. As numerically illustrated in Section 2, under this regime, our model (1) leads to the same dynamics of the moments as in Ronce and Kirkpatrick 2001. Consequently, this equilibrium analysis corresponds to the one made in Ronce and Kirkpatrick 2001 (in the limit of vanishing variance where their system of four equations converges to the system (16)), that revealed the existence of two types of equilibria that could coexist:

- symmetrical, where both populations are of the same size and equally maladapted to their local habitat (corresponding to a generalist species). The authors derived these equilibria analytically. It is worthy to note that in the small variance regime, they reduce to a single monomorphic equilibrium.

- asymmetrical, where one larger population of locally well adapted individuals acts as a source for the other more poorly adapted smaller population (corresponding to a specialist species). The authors explored numerically this type of equilibrium and derived approximations for low migration rates. One aim of this section is to characterize such equilibria analytically.

The fast/slow analysis done in Section 3 gives us the opportunity to go further in the equilibrium analysis in the small variance regime, as the asymptotic system (16) presents a reduced complexity (three equations instead of four). Moreover, adopting the notation $\rho^* = \frac{N_2^*}{N_1^*} > 0$ and using the polynomial previously defined:

$$P_{z^*}(X) = X^3 - X^2 \left[1 + \frac{g}{m}(z^*+1)^2 - \frac{1}{m}\right] + X \left[1 + \frac{g}{m}(z^*-1)^2 - \frac{1}{m}\right] - 1,$$

the Lemma 1 implies that (16) is equivalent to:

$$\begin{cases} P_{z^*}(\rho^*) = 0, \\ \frac{dz^*}{dt} = 2g\left(\frac{\rho^{*2}-1}{\rho^{*2}+1} - z^*\right), \end{cases} \quad (22)$$

with the constraint $\rho^* > \max\left(1 + \frac{g}{m}(z^*+1)^2 - \frac{1}{m}, 0\right)$ (ie. $N_1^* > 0$). This reduction in the regime of small variance allows us in a second time to derive analytical expressions of every possible equilibrium $(z^*, N_1^*, N_2^*) \in \mathbb{R} \times \mathbb{R}_+^{*2}$ from solving:

$$\left[P_{\frac{\rho^{*2}-1}{\rho^{*2}+1}}(\rho^*) = 0\right] \wedge \left[\rho^* > \max\left(1 + \frac{g}{m}\frac{4\rho^{*4}}{(\rho^*+1)^2} - \frac{1}{m}, 0\right)\right], \quad (23)$$

and next setting:

$$\begin{cases} z^* = \frac{\rho^{*2}-1}{\rho^{*2}+1}, \\ N_1^* = m\left[\rho^* - \left[1 + \frac{g}{m}(z^*+1)^2 - \frac{1}{m}\right]\right], \\ N_2^* = m\left[\frac{1}{\rho^*} - \left[1 + \frac{g}{m}(z^*-1)^2 - \frac{1}{m}\right]\right]. \end{cases} \quad (24)$$

We will show that there exists a unique symmetrical equilibrium, which correspond to the monomorphic one analytically found by Ronce and Kirkpatrick 2001 (in the regime of small variance). We will then show that there can additionally exist a mirrored pair of asymmetrical equilibria uniquely defined, corresponding to the ones found numerically by Ronce and Kirkpatrick 2001.

### 4.1 Equilibrium analysis

The objective of this section is to find the steady states $(z^*, N_1^*, N_2^*)$ of the system (16) that lie in $\mathbb{R} \times \mathbb{R}_+^{*2}$ (or equivalently, solve (23) and set (24)). Henceforth, we will call these $(z^*, N_1^*, N_2^*)$ **equilibria**. The systems (23) and (24) imply that $(z^*, N_1^*, N_2^*) \in \mathbb{R} \times \mathbb{R}_+^{*2}$ is an equilibrium if and only if $\bar{Y}^* = \left(N_1^*, N_2^*, \frac{2g}{m\left[\rho^* + \frac{1}{\rho^*}\right]}\right)$ is a fast equilibrium corresponding to $z^* = \frac{\rho^{*2}-1}{\rho^{*2}+1}$. As a corollary of the Proposition 3.1, we get that the following region of parameters does not allow for any equilibria to exist:

**Corollary 1.** *If* $\left[ [g \geq 1] \wedge \left[ m \geq 2g\left(1 - \sqrt{1 - \frac{1}{g}}\right)\right]\right]$, *then there can exist no equilibria as defined by* (23) *and* (24), *i.e. that leads to* $N_1^* > 0$ *and* $N_2^* > 0$.

**Remark 4.1.** *Although our analysis is not meant to describe extinction, we observe numerically that the system goes to extinction in the region defined in the previous corollary (see Fig. 6).*

From now on and until further notice, we will thus consider $(m, g) \in \mathbb{R}_+^{*\,2}$ such that:

$$[g < 1] \vee \left[ m < 2g\left(1 - \sqrt{1 - \frac{1}{g}}\right)\right].$$

### 4.1.1 Symmetric equilibrium: fixation of a generalist species

**Definition 1.** *We call symmetric equilibrium the* $(z^*, N_1^*, N_2^*) \in \mathbb{R} \times \mathbb{R}_+^{*\,2}$ *solutions of* (23) *and* (24) *where both subpopulations have the same size:* $N_1^* = N_2^* = N^* > 0$.

We first state that there can only exist one viable symmetrical equilibrium:

**Proposition 4.1.** *There exists a single symmetric equilibrium when* $g < 1$, *given by* $(0, 1-g, 1-g)$ *and none when* $g \geq 1$.

*Proof.* Regarding (23): we have $\rho^* = 1$ is a positive root of:

$$P_{z^*=0}(X) = X^3 - (1 + \frac{g-1}{m})X^2 + (1 + \frac{g-1}{m})X - 1,$$

that additionally satisfies:

$$\rho^* > 1 + \frac{g-1}{m} \iff 1 > g.$$

Hence the symmetrical equilibrium is uniquely defined by $(0, 1-g, 1-g)$ (from considering (24)). $\square$

In this case, as 0 is the middle point between the local optimal phenotypes $-1$ in habitat 1 and 1 in habitat 2, each subpopulation is equally maladapted.

**Remark 4.2.** *The existence of this equilibrium (or the associated extinction when it is not viable) was expected, for we consider symmetrical habitats and thus symmetrical dynamics. Therefore, under symmetrical initial conditions, the outcome is necessarily symmetrical.*

### 4.1.2 Asymmetric equilibrium: specialist species

We define as asymmetric equilibrium any solution of (24) in $\mathbb{R} \times \mathbb{R}_+^{*\,2}$ that is not a symmetric equilibrium.

**Remark 4.3.** *One can notice that the system* (23) *is invariant under the transformation* $\rho^* \mapsto \frac{1}{\rho^*}$ *or equivalently* (24) *under* $(z^*, N_1^*, N_2^*) \mapsto (-z^*, N_2^*, N_1^*)$. *Thus, we do not lose in generality if we look for equilibria with* $N_1^* < N_2^*$ *instead of* $N_1^* \neq N_2^*$: *to each asymmetrical equilibrium with* $N_1^* < N_2^*$, *we can associate its mirrored version.*

This section is dedicated to confirm the numerical intuition of Ronce and Kirkpatrick 2001 and show that there exists a range of parameters such that a unique mirrored couple of asymmetrical equilibria exists.

**Proposition 4.2.** *Let $(m, g) \in \mathbb{R}_+^{*\,2}$ be such that:*

$$[1 + 2m < 5g] \wedge \left[m^2 > 4g(m-1)\right]. \tag{25}$$

*Then there exists a single asymmetrical equilibrium $(z^*, N_1^*, N_2^*)$ with $N_1^* < N_2^*$, given by:*

$$\begin{cases} N_1^* = (1-m) + m\rho - 4g\dfrac{\rho^{*4}}{(\rho^{*2}+1)^2}, \\[2mm] N_2^* = (1-m) + \dfrac{m}{\rho^*} - 4g\dfrac{1}{(\rho^{*2}+1)^2}, \\[2mm] z^* = \dfrac{\rho^{*2}-1}{\rho^{*2}+1} \neq 0, \end{cases} \tag{26}$$

*where $\rho^* = \dfrac{y^* + \sqrt{y^{*2}-4}}{2}$ and $y^*\left(= \rho^* + \dfrac{1}{\rho^*}\right)$ is the only root greater than 2 of the polynomial:*

$$S(Y) = Y^3 + \dfrac{(1-4g)}{m}Y^2 - \dfrac{4g}{m}Y + \dfrac{4g}{m}.$$

*Conversely, if the condition (25) is not verified, there can be no asymmetrical equilibria.*

**Remark 4.4.** *For $g > 1, m > 0$, we have the equivalence:*

$$[1 + 2m < 5g] \wedge \left[m^2 > 4g(m-1)\right] \iff \left[m < 2g\left(1 - \sqrt{1 - \dfrac{1}{g}}\right)\right].$$

Fig. 4 summarizes the conditions obtained with Proposition 4.1 and Proposition 4.2. It illustrates the analytical range of parameters where the different types of equilibrium exist when the strength of selection $g$ and the migration rate $m$ vary. In the region where none of the conditions are met, we observe numerically that the system leads to extinction (upper right region). In the intermediate green triangle, the two asymmetrical equilibria coexist with the symmetrical equilibrium.

*Proof of Proposition 4.2.* The first part of the proof is directed to solve the equation given in (23) and consists in two lemmas. The second part of the proof examines the conditions under which such solutions verify the inequality constraint given by (23). It consists in a lemma that involves tedious computations. Consequently, the second part of the proof is left to be consulted in Appendix G.

**First part of the proof.** (23) provides us with a close equation: $P_{\frac{\rho^{*2}-1}{\rho^{*2}+1}}(\rho^*) = 0$. Solving it seems necessary, however, the direct search for solutions of this equation leads to consider a seventh degree polynomial. The first part of the proof consists in two lemmas. We first rely on the symmetry of the system noticed by Remark 4.3 (($z^*, \rho^*$) is solution if and only if $\left(-z^*, \dfrac{1}{\rho^*}\right)$ is too) to reduce the complexity from a seventh degree polynomial to a third degree polynomial $S$:

**Lemma 7.** *Let us define:*

$$S(Y) = Y^3 + \dfrac{(1-4g)}{m}Y^2 - \dfrac{4g}{m}Y + \dfrac{4g}{m}.$$

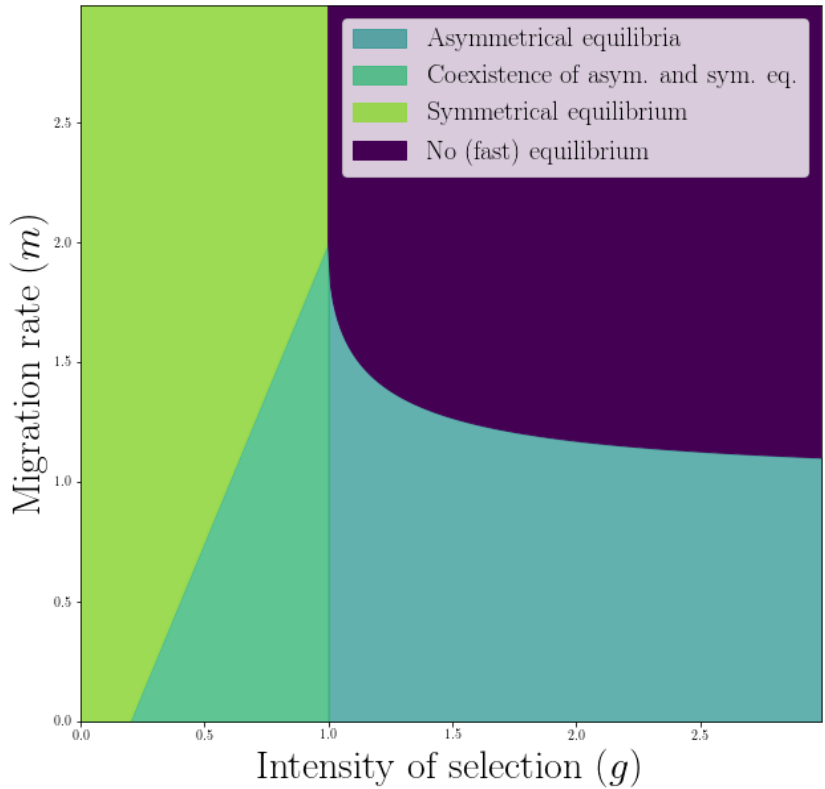

Figure 4: **Regions of existence of the equilibria, according to Proposition 4.1 and Proposition 4.2**. The symmetrical equilibrium is only determined by the intensity of selection, regardless of the migration rate. The asymmetrical equilibria cannot exist for large migration rate ($m > 2$) or small intensity of selection. The limit of the blue region is given by $m = 1$ when $g$ goes to $\infty$. Interestingly enough, at intermediate migration: $m \in [1, 2]$, asymmetrical equilibria only exist for a bounded range of positive $g$: selection cannot be too strong nor too weak. Moreover, see Section 4.2 for stability results about these equilibria to determine which equilibria prevail when both symmetrical and asymmetrical coexist (turquoise triangular region).

Then, we have the following relation for $\rho^* \in \mathbb{R}_+^* \setminus \{1\}$:

$$S\left(\rho^* + \frac{1}{\rho^*}\right) = \frac{(1+\rho^{*2})^2}{(\rho^* - 1)\rho^{*3}} P_{\frac{\rho^{*2}-1}{\rho^{*2}+1}}(\rho^*).$$

As for $\rho^* \in \mathbb{R}_+^* \setminus \{1\}$, $\rho^* + \frac{1}{\rho^*} > 2$, we next look for the number of roots of $S$ greater than 2:

**Lemma 8.** *Let $a > 0, b \in \mathbb{R}$. Let us define $b(a) := \frac{5a}{4} - 2$. Then: if $b \geq b(a)$, $S(Y) = Y^3 + (b-a)Y^2 - aY + a$, has no root greater than 2. If $b < b(a)$, $S$ has a single root greater than 2.*

The successive application of the Lemma 7 and Lemma 8 with:

$$\begin{cases} b = \frac{1}{m}, \\ a = \frac{4g}{m} > 0, \end{cases}$$

yields that there exists a unique solution to (23) if and only if $1 + 2m < 5g$, and therefore to (24) in $\mathbb{R} \times \mathbb{R}_+^{*2}$ which is exactly (26). Proving the two lemmas concludes the first part of the proof.

*Proof of Lemma 7.* Let us consider $\rho^* \in \mathbb{R}^* \setminus \{1\}$. Then we have:

$$\frac{(1+\rho^{*2})^2}{(\rho^* - 1)\rho^{*3}} P_{\frac{\rho^{*2}-1}{\rho^{*2}+1}}(\rho^*) = \frac{2-4g}{m} + \frac{(3m-4g)}{m}\left(\rho^* + \frac{1}{\rho^*}\right) + \frac{(1-4g)}{m}\left(\rho^{*2} + \frac{1}{\rho^{*2}}\right) + \rho^{*3} + \frac{1}{\rho^{*3}}$$

$$= \frac{2-4g}{m} + \frac{(3m-4g)}{m}\left(\rho^* + \frac{1}{\rho^*}\right) + \frac{(1-4g)}{m}\left(\rho^{*2} + \frac{1}{\rho^{*2}}\right) + \rho^{*3} + \frac{1}{\rho^{*3}}.$$

Since:

$$\rho^{*2} + \frac{1}{\rho^{*2}} = \left(\rho^* + \frac{1}{\rho^*}\right)^2 - 2,$$

$$\rho^{*3} + \frac{1}{\rho^{*3}} = \left(\rho^* + \frac{1}{\rho^*}\right)^3 - 3\left(\rho^* + \frac{1}{\rho^*}\right),$$

we have:

$$\frac{(1+\rho^{*2})^2}{(\rho^* - 1)\rho^{*3}} P_{\frac{\rho^{*2}-1}{\rho^{*2}+1}}(\rho^*) = \left(\rho^* + \frac{1}{\rho^*}\right)^3 + \frac{1-4g}{m}\left(\rho^* + \frac{1}{\rho^*}\right)^2 - \frac{4g}{m}\left(\rho^* + \frac{1}{\rho^*}\right) + \frac{4g}{m}$$

$$= S\left(\rho^* + \frac{1}{\rho^*}\right).$$

$\square$

*Proof of Lemma 8.* As $S(0) = a > 0$ and since $S$ goes to $-\infty$ in $-\infty$, $S$ has always a negative root.

Thereby, the case that we take interest in is included within the case where all three roots $Z_1, Z_2, Z_3$ of $S$ are real. Furthermore, we have the following relations:

$$\begin{cases} Z_1 Z_2 Z_3 = -a < 0, \\ Z_1 Z_2 + Z_2 Z_3 + Z_3 Z_1 = -a < 0. \end{cases}$$

From the first relation, we deduce that $S$ has an even number of positive roots, so either 0 or 2. The second relation leads to a contradiction if all roots are negative. Thus $S$ has necessarily two positive roots and one negative.

Moreover, we have:
$$\frac{1}{Z_1} + \frac{1}{Z_2} + \frac{1}{Z_3} = \frac{Z_1 Z_2 + Z_2 Z_3 + Z_3 Z_1}{Z_1 Z_2 Z_3} = 1.$$

Without loss of generality, let us assume that $Z_3 < 0$. If the remaining two positive roots were greater than 2, then we would get:
$$1 < \frac{1}{Z_1} + \frac{1}{Z_2} \leq \frac{1}{2} + \frac{1}{2} = 1$$

which is a contradiction. Hence at most one is greater than or equal to 2.

The only fact that is left to prove is that such a root exists. Let $S_a(X) = X^3 + (b(a) - a)X^2 - aX + a$. Under the choice of $b(a)$, we can verify that $S_a(2) = 0$. Consequently, the following holds:
$$b < b(a) \iff S(2) < S_a(2) = 0.$$

Therefore, because $S$ goes to $+\infty$ in $+\infty$, if $b > b(a)$, $S$ has an even number of roots greater than 2. Thereby, from the previous part of the proof, in that case, $S$ do not have any roots greater than 2. If $b = b(a)$, 2 is the only root of $S$ greater than or equal to 2. If $b < b(a)$, $S$ has at least one root strictly greater than 2. This root is unique by the argument above (which was independent of $b$).

$\square$

**Second part of the proof.** The second part of the proof is dedicated to show that for all $(m, g) \in \mathbb{R}_+^{*\,2}$ verifying (25), the solution $\rho^* > 0$ that we found in the first part of the proof verifies the constraint given in (23). It consists in the following lemma, that is obtained after tedious calculations, so the proof is left to be consulted in Appendix G.

**Lemma 9.** *Let $(m, g) \in \mathbb{R}_+^{*\,2}$ verifying (25), and $\rho^* > 0$ be the unique solution of the equation $\left[P_{\frac{\rho^{*2}-1}{\rho^{*2}+1}}(\rho^*) = 0\right]$. Then:*
$$\rho^* > 1 + \frac{g}{m} \frac{4\rho^{*4}}{(\rho^* + 1)^2} - \frac{1}{m}.$$

Consequently: for $(g, m) \in \mathbb{R}_+^{*\,2}$ such that $1 + 2m < 5g$ and $m^2 > 4g(m - 1)$, $\rho^*$ defined in Proposition 4.2 defines an equilibrium with positive subpopulation sizes.

Conversely: if (25) is not met, either $1 + 2m > 5g$, in which case no asymmetrical equilibrium can exist from Lemma 7 and Lemma 8, or $m^2 < 4g(m - 1)$ (which implies that $g > 1$), in which case Remark 4.4 and Corollary 1 implies that no equilibrium can exist. $\square$

## 4.2 Stability analysis

In this subsection, we examine the stability of the equilibria of the system (22) that we described previously.

**Proposition 4.3.** Let $(z^*, N_1^*, N_2^*) \in \mathbb{R} \times \mathbb{R}_+^{*\,2}$ be an equilibrium and $\rho^* = \frac{N_2^*}{N_1^*}$. Then the equilibrium is locally stable (respectively unstable) if:

$$\frac{4\rho^*}{(\rho^{*2}+1)^2} \times \frac{1}{P'_{z^*}(\rho^*)} \times \frac{2g}{m}\left[z^*\left(\rho^* - \rho^{*2}\right) - \left(\rho^* + \rho^{*2}\right)\right] + 1 > 0 \quad (resp. < 0).$$

*Proof.* If $(z^*, N_1^*, N_2^*) \in \mathbb{R} \times \mathbb{R}_+^{*\,2}$ is an equilibrium and $\rho^* = \frac{N_2^*}{N_1^*}$, then $(N_1^*, N_2^*)$ is a fast equilibrium associated to $z^*$ (Lemma 1), which implies that $P_{z^*}$ has a single positive root (without multiplicity) that is $\rho^*$ (Lemma 2). Hence $\rho^*$ cannot be a double root of $P_{z^*}$, which yields: $P'_{z^*}(\rho^*) \neq 0$.

(22) implies that the local stability of the equilibria can be examined by the following system:

$$\begin{cases} \mathcal{G}(z^*, \rho^*) := P_{z^*}(\rho^*) = 0, \\ \rho^* > \left[1 + \frac{g}{m}(z^*+1)^2 - \frac{1}{m}\right], \\ \frac{dz^*}{dt} = \mathcal{F}(z^*, \rho^*) := 2g\left(\frac{\rho^{*2}-1}{\rho^{*2}+1} - z^*\right). \end{cases}$$

As $\partial_\rho \mathcal{G}(z^*, \rho^*) = P'_{z^*}(\rho^*) \neq 0$, we apply the implicit function theorem to get $U$ a open neighbourhood of $z^*$ and a smooth function $\rho : U \to \mathbb{R}_+^*$ such that:

$$\forall z \in U, \mathcal{G}(z, \rho(z)) = 0.$$

For $z \in U$, we define $f : U \to \mathbb{R}$, $z \mapsto \mathcal{F}(z, \rho(z))$. Hence, $(z^*, N_1^*, N_2^*)$ is locally stable (resp. unstable) if :

$$f'(z^*) = \nabla \mathcal{F}(z^*, \rho^*) \cdot \begin{pmatrix} 1 \\ \frac{d\rho}{dz}(z^*) \end{pmatrix} = \partial_\rho \mathcal{F}(z^*, \rho^*)\left[-(\partial_\rho \mathcal{G}(z^*, \rho^*))^{-1} \partial_z \mathcal{G}(z^*, \rho^*)\right] - 2g < 0 \quad (resp. > 0).$$

Since we have:

$$\partial_\rho \mathcal{F}(z^*, \rho^*) = 2g\frac{4\rho^{*2}}{(\rho^*+1)^2}, \quad (\partial_\rho \mathcal{G}(z^*, \rho^*))^{-1} = \frac{1}{P'_{z^*}(\rho^*)},$$

and

$$\partial_z \mathcal{G}(z^*, \rho^*) = -2\frac{g}{m}(z^*+1)\rho^{*2} + 2\frac{g}{m}(z^*-1)\rho^*,$$

the considered equilibrium is locally stable (reps. unstable) if:

$$\frac{4\rho^*}{(\rho^{*2}+1)^2} \times \frac{1}{P'_{z^*}(\rho^*)} \times \frac{2g}{m}\left[z^*\left(\rho^* - \rho^{*2}\right) - \left(\rho^* + \rho^{*2}\right)\right] + 1 > 0 \quad (resp. < 0).$$

$\square$

**Corollary 2.** *The symmetrical equilibrium $z^* = 0, \rho^* = 1$ is locally stable (resp. unstable) if $5g < 1 + 2m$ (resp. $5g > 1 + 2m$) (ie, when it is alone).*

*Proof.* If $z^* = 0$ and $\rho^* = 1$, we have:

$$\frac{4\rho^*}{(\rho^{*2}+1)^2} \times \frac{1}{P'_{z^*}(\rho^*)} \times \frac{2g}{m}\left[z^*\left(\rho^* - \rho^{*2}\right) - \left(\rho^* + \rho^{*2}\right)\right] + 1$$

$$= \frac{1}{3 - 2\left(1 + \frac{g}{m} - \frac{1}{m}\right) + \left(1 + \frac{g}{m} - \frac{1}{m}\right)} \times \frac{-4g}{m} + 1 = \frac{1 + 2m - 5g}{1 + 2m - g}.$$

We recall that for the symmetrical equilibrium to exist, we need: $g < 1$, which imply: $g < 1 + 2m$. Hence the result. $\square$

Analytical derivations are more tedious for asymmetrical equilibria. However, when $1 + 2m > g$, we showed that $P_{z^*}$ has a single (without multiplicity) positive root $\rho(z^*)$ for all $z^* \in [-1, 1]$ (Proposition 3.2). The function $\rho : [-1, 1] \to \mathbb{R}_+^*, z \mapsto \rho(z)$ is therefore smooth (where $\rho(z)$ designates the single positive root of $P_z$). Thus, we can globally define the smooth function $f$ similarly as in Proposition 4.3 on $]-1, 1[$:

$$f : \begin{cases} ]-1, 1[ \to \mathbb{R} \\ z \mapsto 2g \left( \frac{\rho(z)^2 - 1}{\rho(z)^2 + 1} - z \right), \end{cases}$$

That leads to the following result:

**Corollary 3.** *Let $5g > 1 + 2m > g$. Then the asymmetrical equilibria are locally stable.*

*Proof.* Let $(z^*, N_1^*, N_2^*)$ be an asymmetrical equilibrium. We recall that $z^* = \frac{\rho^{*2} - 1}{\rho^{*2} + 1} \in ]-1, 1[$. From the previous corollary, the symmetric equilibrium is locally unstable, i.e.:

$$f'(0) > 0.$$

Moreover, from Proposition 3.2, $P_{z^*=1}$ has a single positive root, and we can extend $f$ in 1 by continuity and calculate :

$$f(1) = 2g \left( \frac{\rho^2(1) - 1}{\rho^2(1) + 1} - 1 \right) = -\frac{4g}{\rho^2(1) + 1} < 0.$$

Since 0 and $z^*$ are the only zeros of $f$ on $[0, 1]$ (from the uniqueness of the mirrored couple of asymmetric equilibria) and $f'(0) > 0$, $f$ is positive on $]0, z^*[$ ane negative on $]z^*, 1]$. Hence, the asymmetrical equilibria are locally stable. □

To illustrate the diversity of cases in both the number of equilibria and their stability, we display in Fig. 5 the graph of the function $f$ defined above as a function of the dominant trait $z$ when $g = 1.5$ and $m$ takes the following values :

1. $m = 0.02$. There are multiple branches near the origin (yellow curve), as the function $f$ is multi-valued. Indeed, we are in the case where: $1 + 2m < g$. Therefore, for $z^*$ near 0, there is three distinct positive roots for $P_{z^*}$ (from Proposition 3.2), which leads to non-viable fast equilibria (from Lemma 2). Therefore, if the initial dominant trait is near 0, the system will go to extinction.

2. $m = 0.25$, so that the equality $1 + 2m = g$ holds, which is the limit case of the folding near the origin.

3. $m = 1$. For each value of the dominant trait $z^*$, there is only one root to $P_{z^*}$. There are three equilibria, an unstable symmetric and two stable asymmetric equilibria ($1 + 2m < 5g$).

4. $m = 3.25$, so that the equality $1 + 2m = 5g$ holds. This displays the limit of existence of the asymmetrical equilibria (see Proposition 4.2). The three equilibria are merging and exchanging stability.

5. $m = 5$. As $m$ grows further, the asymmetric equilibria do not exist anymore. Therefore, only the symmetric one is left and is stable.

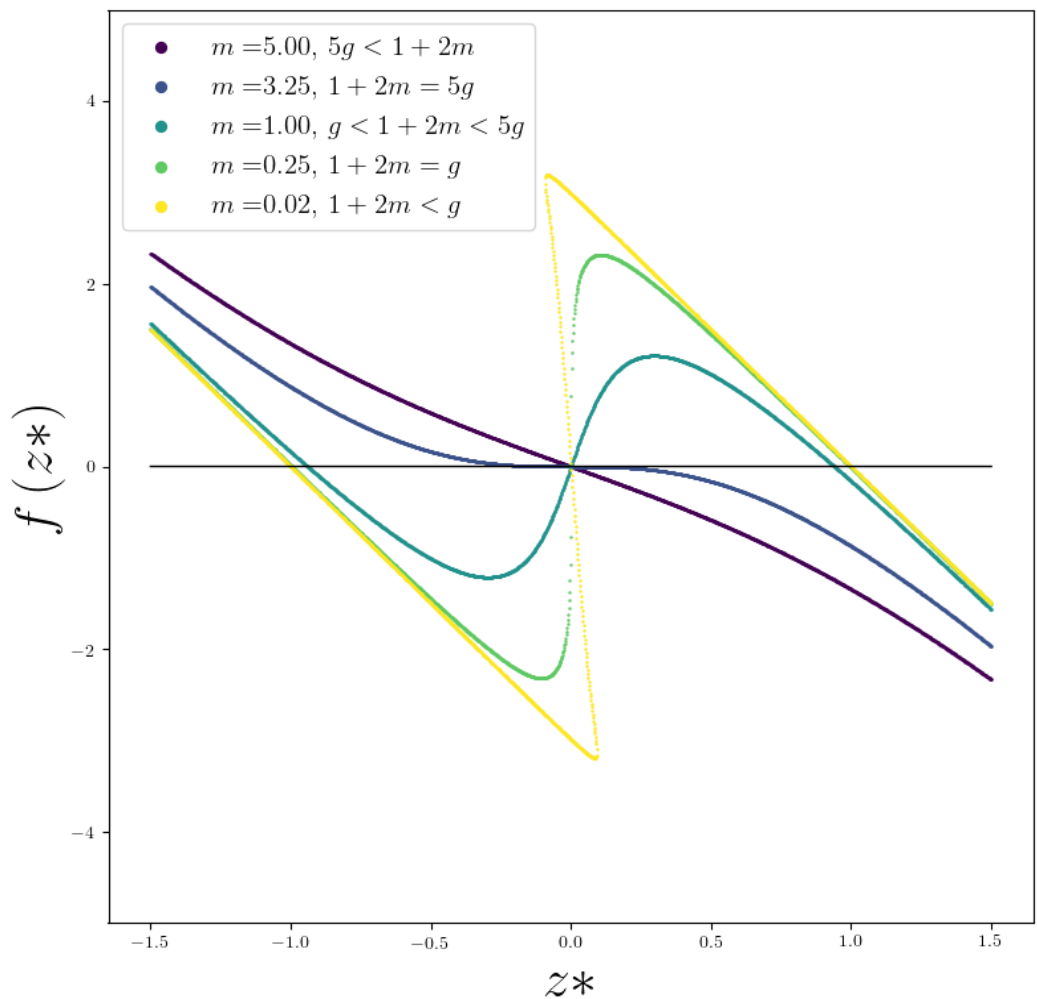

Figure 5: **Graph of the function $f$ for $g = 1.5$ and $m \in \{0.02, 0.25, 1, 3.25, 5\}$.** The dominant traits $z^*$ of the different equilibria are located where the curve crosses the horizontal line ($f(z^*) = 0$). An equilibrium is locally stable if the slope of $f$ at the point of equilibrium is negative. For decreasing values of $m$ (dark to light colors), at first, only the symmetric equilibrium exists and is stable (see also Corollary 2). Then, the asymmetrical equilibria emerge (in the parameter region indicated in Proposition 4.2) and are bistable (see also Corollary 3), while the symmetric equilibrium becomes unstable. For small values of $m$, the curve folds near the origin, as for $z^*$ near 0, $P_{z^*}$ has three distinct positive roots (Proposition 3.2). For those $z^*$, the fast equilibria are all non viable (Lemma 2): numerically, the system goes to extinction if the initial dominant trait is near 0.

# 5  Discussion

**Contributions**  In this paper, we have studied the evolutionary dynamics of a complex trait under stabilizing selection in a heterogeneous environment in a sexually reproducing population. To model the process of inheritance of this trait, we have used a mixing sexual reproduction operator according to the infinitesimal model (Fisher 1919; Bulmer 1971; Barton, Etheridge, and Véber 2017), assuming that the segregational variance is constant and independent of the families. We have set our analysis in a regime of small variance of segregation, aligning with a framework developed by Diekmann et al. 2005; Perthame and Barles 2008 and recurrently used with the infinitesimal model "Equilibria of quantitative genetics models beyond the Gaussian approximation I: Maladaptation to a changing environment"; Calvez, Garnier, and Patout 2019. By doing so, we showed two types of result. First, we compared the system of moments derived from our model in the limit of small variance with a seminal work in quantitative genetics (Ronce and Kirkpatrick 2001), showing their equivalence in that limit, while bypassing any prior normality assumption on the trait distributions. Next, we showed that this small variance regime discriminates two time scales, allowing to perform a slow-fast analysis, which reduces the complexity of our system in the asymptotic limit. Thus, we were able to fully derive analytically its equilibria thanks to algebraic arguments of symmetry reflecting the symmetrical habitats. The theoretical outcomes of our model are shown in the upper panel of Fig. 6. They are to be compared to numerical outcomes shown in the lower panel, where the same colours indicate the same types of equilibria. For the numerical analysis, for each couple of parameters $(m, g)$, the initial state is the same: both local distributions are normal of same mean (0.2) and same variance $\varepsilon^2 = 2.5 \times 10^{-3}$. The initial state is taken as monomorphic so that it falls within the scope of the slow-fast analysis. Moreover, the color yellow is attributed to simulations whose final state does not meet the small segregational variance regime analysis prediction, which in particular states that the distribution of trait in the meta-population has a variance of order $\varepsilon^2$ (see (12) and recall that the population is monomorphic (Section 3)). In the two simulations that present the color yellow, the variance in trait in the meta population is of approximately $3\,\varepsilon^2$, which exceeds the chosen threshold $(2\,\varepsilon^2)$. The detailed setting and scoring of the simulations involved in the lower panel of Fig. 6 are available in Appendix H.

One can notice that the justification of the validity of the Gaussian approximation of local trait distributions in the regime of small variance (see Section 1 and "Equilibria of quantitative genetics models beyond the Gaussian approximation I: Maladaptation to a changing environment") and most of the slow-fast analysis (Section 3) are robust when introducing asymmetries in our model, or changing the selection functions. However, we stress that our analytical derivation of the equilibria in the asymptotic limit uses specific arguments that rely crucially on the symmetries between habitats in our model (see Remark 4.3 and Proposition 4.2).

**Robustness with regard to dimorphic initial state.**  The theoretical outcomes given in Fig. 6 are in particular a consequence of the reduction of system due to the slow-fast theorem, which applies provided that the initial state is close enough from a fast equilibria from the slow manifold (see Theorem 3.1). Those fast equilibria are monomorphic. A natural question would be to ask to what extent those results apply for an initial state that is dimorphic. This would model for example two initially isolated subpopulations, locally adapted, that are suddenly being connected. Here we give a numerical taste of what a more complete answer could look like. We display Fig. 7 using the same methodology and scoring than for the lower panel of Fig. 6, the only difference being the initial state, now constituted by two locally adapted subpopulations, slightly asymmetrical in size (see Appendix H for details). To get a sense of what could occur in the regime of vanishing variance, we choose to display the results for two values of $\varepsilon^2$: $\varepsilon^2 = 2.5 \times 10^{-3}$ (upper panel) and $\varepsilon^2 = 6.25 \times 10^{-4}$ (lower panel). Both panels of Fig. 7 and the lower panel of Fig. 6 are globally quite similar, except for the yellow region that is much wider in both panels of Fig. 7. Particularly, there is a net trend for strong selection and small migration. That is expected, because the initial state of the simulations involved in both panels of Fig. 7 is presumably far from the conditions asked by Theorem 3.1. These simulations suggest that, in this particular range of parameters, the fast relaxation to a monomorphic state, that is central in Theorem 3.1, breaks down and dimorphism is maintained. However, we can note that this yellow region decreases for decreasing values of $\varepsilon^2$ (difference between upper and lower panel of Fig. 7). That suggests that our analysis remains quite robust to dimorphic initial states in the limit of vanishing variance.

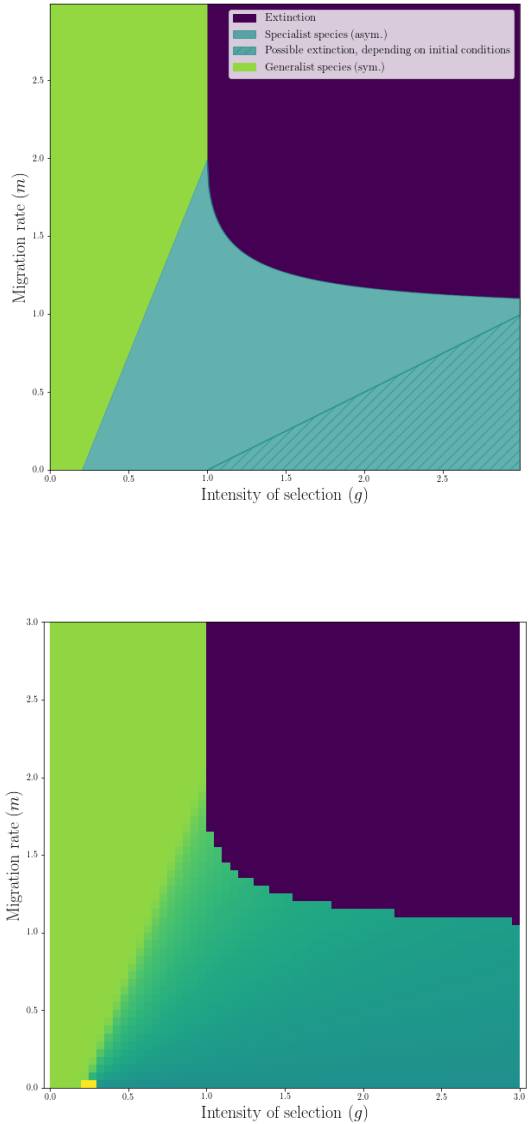

Figure 6: **Summary of the different theoretical (upper panel, in the limit of vanishing segregational variance) and numerical (lower panel, $\varepsilon^2 = 2.5 \times 10^{-3}$) outcomes of our model when selection ($g$) and migration ($m$) are varying**. The same colors represents the same outcomes in both figures. Fig. 4 complemented by the stability analysis (see Section 4.2) gives the upper figure. In the dashed region, the system goes to one of the asymmetrical equilibrium, except if the initial conditions are too symmetrical (the system goes then numerically to extinction, typically due to the folding near $z^* = 0$ of the yellow curve in Fig. 5). For the lower figure, all simulations share the same initial state: the meta-population is monomorphic and asymmetrical as local distributions are both normal with same mean (0.2) and same variance ($2\varepsilon^2$). Hence, the potential extinction in the dashed region does not occur and the numerical analyis falls within the scope of the slow-fast analysis (Section 3). The color yellow is attributed to simulations whose final state does not meet the small segregational variance regime analysis prediction, which, in particular, states that the distribution of trait in the meta-population has a variance of order $\varepsilon^2$ (see (12) and recall that the population is monomorphic (Section 3)). For more details on the simulations and their scoring resulting in the lower panel figure, see Appendix H.

**Comparison with asexual studies.** In Section 4, we found that bistable asymmetrical equilibria can exist in our system (Proposition 4.2, Corollary 3). That is a strong difference with the findings of Mirrahimi 2017 and Mirrahimi and Gandon 2020: with a similar mesoscopic model but using an asexual reproduction operator with frequent mutations of small effects, they find that symmetrical habitats lead to a single stable symmetrical equilibrium, either monomorphic or dimorphic. In particular, if migration is small enough compared to selection, each subpopulation adapts to their habitats and dimorphism occurs at the metapopulation scale. In our case, the mixing effect of the infinitesimal operator of sexual reproduction does not allow for such a local adaptation to occur in the limit of small variance. In Section 3, we showed that it forces monomorphism quickly and the only option to adapt to strong forces of selection is an asymmetrical equilibrium (Proposition 4.2, Fig. 6) that describes a source sink scenario. One population is adapted to its habitat, and the other is essentially composed by poorly adapted migrants ; the choice of which depends on the initial conditions.

Our findings share notable similarities with some in Débarre, Ronce, and Gandon 2013, which conducts a hybrid analysis on asexual populations with tools of adaptive dynamics applied to quantitative genetics equations. Particularly, under gradual evolution (when mutations are rare and of small effects), they state that asymmetrical equilibria can be reached if the population is initially monomorphic, under a similar range of migration and selection parameters as indicated by our analysis. To solve for them, they assume that the distributions of traits around each peak found using adaptive dynamics is Gaussian, of constant variance related to the mutational variance (which is small by hypotheses). That is similar to the framework that naturally arise from the hypotheses of our model, should the mutational variance be replaced by the segregational variance. Consequently, we suggest that the asymmetrical equilibria found in Débarre, Ronce, and Gandon 2013 should have the same coordinates as the ones found in our analysis. However, there is a substantial difference in the dynamics leading to those equilibria. Even with an initially dimorphic meta-population, our hypotheses on sexual reproduction typically strains toward monomorphism. With the same initial state, Débarre, Ronce, and Gandon 2013 indicate that dimorphism is typically maintained in the range of parameters where asymmetrical equilibria exist.

**Gaussian assumption.** In our study, we showed that when the segregational variance is small compared to how differently the two habitats select, the local trait distributions can be approximated by normal distributions (Section 2). Hence, asymptotically, in the regime of small variance, the findings of our model are equivalent to Ronce and Kirkpatrick 2001, which relies on a Gaussian assumption of local trait distributions. This link of equivalence relies on the hypothesis that the genetic (and phenotypic) variance is constant, which we interpreted in our model to be twice the segregational variance in the limit of vanishing variance. Furthermore, together with the last paragraph, our study gives some elements of explanation to why the findings of Ronce and Kirkpatrick 2001 are structurally different from Mirrahimi 2017 and Mirrahimi and Gandon 2020, and closer to Débarre, Ronce, and Gandon 2013.

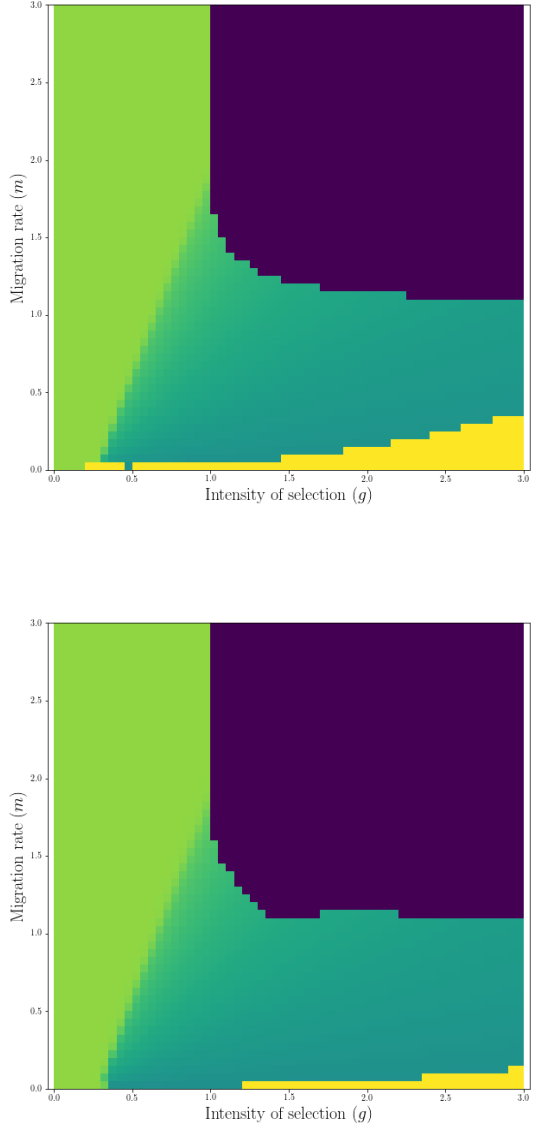

Figure 7: **Numerical outcomes with initial dimorphic state (locally adapted subpopulations), for $\varepsilon^2 = 2.5 \times 10^{-3}$ (upper panel) and $\varepsilon^2 = 6.25 \times 10^{-4}$ (lower panel).** The colors for both figures results from the same scoring scheme as for Fig. 6 (see Appendix H for details). The results are quite similar as Fig. 6, except for the yellow region. In the upper panel ($\varepsilon^2 = 2.5 \times 10^{-3}$), the yellow region is wider than when the initial state is monomorphic (Fig. 6), increasingly so for stronger selection. That highlights the numerical cases where the population ends up dimorphic as the species adapts locally to each deme's optimum, for strong selection and small migration. This is expected as the fast convergence toward a monomorphism state induced by Theorem 3.1 in the limit of vanishing variance of segregation is likeky to break down, as the initial state is far from the slow manifold and the segregational variance is small but not zero. However, this yellow region decreases as the value of $\varepsilon^2$ decreases, as indicated by the lower panel ($\varepsilon^2 = 6.25 \times 10^{-4}$). That suggests that our analysis remains quite robust to dimorphic initial states in the limit of vanishing variance.

**Constant segregational variance in a heterogeneous environment** Our model relies on using the infinitesimal model with a constant segregational variance, independent of the mates deme. That is a strong assumption. However, in the perspective of linking the present study to population genetics approaches, one can question the limits of such a modelling assumption with regard to a Mendelian interpretation of this model. A future work is planned to examine it through conducting individual based simulations.

# Acknowledgements

The author thanks Vincent Calvez and Sepideh Mirrahimi for supervising this project and Sarah P. Otto for precise and helpful comments. The author also thanks Ophélie Ronce, Florence Débarre, Amandine Véber, Alison Etheridge and Florian Patout for enlightening conversations. This project has received funding from the European Research Council (ERC) under the European Union's Horizon 2020 research and innovation programm (grant agreement No 639638).

# Appendices

## A System of moments derived from our model

Here, we derive the system of moments (2) from (1). In the preliminary computations, we will omit the time and deme dependency for the sake of clarity. We will then denote $\boldsymbol{n}$ the trait distribution density, $\boldsymbol{N}$ the size of the population, $\overline{\boldsymbol{z}}$ the mean trait, $\boldsymbol{\sigma}^2$ the mean variance, $\boldsymbol{\psi}^3$ the third central moment and $\boldsymbol{\theta}$ the optimal phenotype.

**Preliminary integration of the selection term.** We have:

$$\int_{\mathbb{R}} (\boldsymbol{z}-\boldsymbol{\theta})^2 \boldsymbol{n}(\boldsymbol{z}) d\boldsymbol{z} = \int_{\mathbb{R}} \left[ (\boldsymbol{z}-\overline{\boldsymbol{z}})^2 + (\overline{\boldsymbol{z}}-\boldsymbol{\theta})^2 + 2(\boldsymbol{z}-\overline{\boldsymbol{z}})(\overline{\boldsymbol{z}}-\boldsymbol{\theta}) \right] \boldsymbol{n}(\boldsymbol{z}) \, d\boldsymbol{z}$$
$$= \boldsymbol{\sigma}^2 \boldsymbol{N} + (\overline{\boldsymbol{z}}-\boldsymbol{\theta})^2 \boldsymbol{N},$$

and:

$$\int_{\mathbb{R}} (z-\overline{z})(z-\theta)^2 n dz = \int_{\mathbb{R}} \left[(z-\overline{z})^3 + (z-\overline{z})(\overline{z}-\theta)^2 + 2(z-\overline{z})^2(\overline{z}-\theta)\right] n(z)\,dz$$
$$= 2\sigma^2(\overline{z}-\theta)N + \psi N.$$

**Size of the subpopulations.** Recalling that $N_i(t) = \int_{\mathbb{R}} n_i(t,z)\,dz$, we get from the preliminary computations by integrating (1):

$$\frac{dN_i}{dt} = \int_{\mathbb{R}} \frac{\partial n_i}{\partial t}(t,z)dz$$
$$= \int_{\mathbb{R}} r\mathcal{B}_\sigma(n_i)(t,z) - g(z-\theta_i)^2 n_i(t,z) - \kappa N_i(t) n_i(t,z) + m\left(n(t,z) - n(t,z)\right)\,dz$$
$$= \left[r - \kappa N_i(t) - g(\overline{z}_i(t) - \theta_i)^2 - g\sigma_i^2\right] N_i(t) + m(N_j(t) - N_i(t)).$$

**Local mean trait.** Recalling that $\overline{z}_i(t) = \frac{1}{N_i(t)} \int_{\mathbb{R}} z\, n_i(t,z)\,dz$, we have, thanks to the preliminary computations:

$$\frac{d\overline{z}_i}{dt} = \frac{1}{N_i} \int_{\mathbb{R}} z \frac{\partial n_i}{\partial t}(t,z)dz - \frac{1}{N_i^2}\frac{dN_i}{dt}\int_{\mathbb{R}} z n_i(t,z)dz$$
$$= \frac{1}{N_i} \int_{\mathbb{R}} (z - \overline{z}_i) \frac{\partial n_i}{\partial t}(t,z)dz$$
$$= \frac{1}{N_i} \int_{\mathbb{R}} (z - \overline{z}_i)\left[-g(z-\theta_i)^2 n_i(t,z) + m(n_j(t,z) - n_i(t,z))\right]dz$$
$$= 2g\sigma_i^2(\theta_i - \overline{z}_i) - g\psi_i^3 + m\frac{N_j}{N_i}(\overline{z}_j - \overline{z}_i).$$

# B Equilibria of a dynamical system under the infinitesimal model of reproduction with random mating only

In this subsection, we show that (7) admits any Gaussian of variance $\varepsilon^2$ as equilibrium. That is equivalent to state that:

**Proposition B.1.** *For $\mu \in \mathbb{R}$, the Gaussian distribution $G_{\mu,\varepsilon^2}$ of mean $\mu$ and variance $\varepsilon^2$ is a fixed point of the operator $\mathcal{B}_\varepsilon$, namely:*

$$\mathcal{B}_\varepsilon(G_{\mu,\varepsilon^2}) = G_{\mu,\varepsilon^2}.$$

*Proof.* We can first notice that $\mathcal{B}_\varepsilon$ can be written using a double convolution product:

**Lemma 10.** *For $f \in \mathcal{L}_1(\mathbb{R})$, $\int_{\mathbb{R}} f \neq 0$, we have:*

$$\mathcal{B}_\varepsilon(f) = \frac{4}{\int_{\mathbb{R}} f(z')\,dz'} G_{0,\frac{\varepsilon^2}{2}} * F * F,$$

*where $F : z \mapsto f(2z)$.*

*Proof of Lemma 10.* For $f \in \mathcal{L}_1(\mathbb{R})$, $\int_{\mathbb{R}} f \neq 0$, a straight-forward computation yields:

$$\mathcal{B}_\varepsilon(f)(z) = \frac{1}{\sqrt{\pi}\varepsilon} \iint_{\mathbb{R}^2} \exp\left[\frac{-(z - \frac{z_1+z_2}{2})^2}{\varepsilon^2}\right] \frac{f(z_1)f(z_2)}{\int_{\mathbb{R}} f(z')\, dz'} dz_1 dz_2$$

$$= \frac{1}{\int_{\mathbb{R}} f(z')\, dz'} \int_{\mathbb{R}} \int_{\mathbb{R}} G_{0,\frac{\varepsilon^2}{2}}\left((z - \frac{z_1}{2}) - \frac{z_2}{2}\right) F(\frac{z_2}{2})\, dz_2\, F(\frac{z_1}{2})\, dz_1$$

$$= \frac{2}{\int_{\mathbb{R}} f(z')\, dz'} \int_{\mathbb{R}} G_{0,\frac{\varepsilon^2}{2}} * F(z - \frac{z_1}{2}) F(\frac{z_1}{2})\, dz_1$$

$$= \frac{4}{\int_{\mathbb{R}} f(z')\, dz'} G_{0,\frac{\varepsilon^2}{2}} * F * F(z).$$

$\square$

If $f = G_{\mu,\varepsilon^2}$, then we find $F = \frac{1}{2} \times G_{\frac{\mu}{2}, \frac{\varepsilon^2}{4}}$. Besides, as the convolution product of two Gaussian kernels $G_{\mu_1,\sigma_1^2}$ and $G_{\mu_2,\sigma_2^2}$ is the Gaussian kernel $G_{\mu_1+\mu_2,\sigma_1^2+\sigma_2^2}$, Proposition B.1 is a corollary of the previous lemma. $\square$

## C Formal expansion within the exponential formalism for $n_\varepsilon$

In this subsection, we will remove the deme dependency for the sake of clarity. To formally derive (9), let us consider the following formal expansion of $U_\varepsilon$ with regard to successive orders of $\varepsilon^2$:

$$U_\varepsilon = u_0 + \varepsilon^2 u_\varepsilon.$$

The aim is to characterize $u_0$ thanks to the behaviour of the reproduction term when $\varepsilon \ll 1$, which we expect neither to diverge nor to vanish:

$$\frac{\mathcal{B}_\varepsilon(n_\varepsilon)}{n_\varepsilon}(z) =$$

$$\frac{1}{\sqrt{\pi}\varepsilon} \iint_{\mathbb{R}^2} \frac{\exp\left[\frac{1}{\varepsilon^2}\left[-\left[z - \frac{z_1+z_2}{2}\right]^2 + u_0(z) - u_0(z_1) - u_0(z_2)\right]\right] \exp\left[u_\varepsilon(z) - u_\varepsilon(z_1) - u_\varepsilon(z_2)\right] dz_1 dz_2}{\int_{\mathbb{R}} \exp\left[-\frac{u_0(z')}{\varepsilon^2} - u(z')\right] dz'}$$

Then, we have several considerations to make. First, if we assume that $u_0$ reaches its minimum at a non degenerate point, then the following modified expression of the denominator:

$$\frac{1}{\sqrt{\pi}\varepsilon} \int_{\mathbb{R}} \exp\left[-\frac{1}{\varepsilon^2}\left[u_0(z') - \min u_0\right] - u(z')\right] dz',$$

will have its integrand concentrate around the minimum of $u_0$ and will converge as $\varepsilon \ll 1$. Therefore it is relevant to introduce this minimum both at the numerator and the denominator.

Then, since we expect the numerator not to diverge nor to vanish uniformly as $\varepsilon \ll 1$, we need that:

$$\forall z \in \mathbb{R}, \max_{(z_1,z_2)} \left[-\left(z - \frac{z_1+z_2}{2}\right)^2 + u_0(z) - u_0(z_1) - u_0(z_2) + \min u_0\right] = 0. \tag{27}$$

As shown in "Equilibria of quantitative genetics models beyond the Gaussian approximation I: Maladaptation to a changing environment", thanks to some convexity arguments, this leads necessarily to choose $u_0$ as a quadratic function in $z$, hence its decomposition:

$$u_0(z) = b + \frac{(z-z^*)^2}{2},$$

where $z^*$ is realizing the minimum of $u_0$.

Finally, let us explain why we set $b = 0$. According to the Laplace method of integration, we have that:

$$\varepsilon^2 \log\left[\int_\mathbb{R} \exp\left[-\frac{U_\varepsilon(z)}{\varepsilon^2}\right] dz\right] \underset{\varepsilon \to 0}{\to} -b,$$

which yields:

$$N_\varepsilon = \frac{1}{\sqrt{2\pi}\varepsilon} \int_\mathbb{R} \exp\left[-\frac{U_\varepsilon(z)}{\varepsilon^2}\right] dz \underset{\varepsilon \to 0}{\approx} \frac{1}{\sqrt{2\pi}\varepsilon} \exp\left[-\frac{b}{\varepsilon^2}\right].$$

So either $b = 0$, either there is extinction or explosion of the population size. That yields (9).

## D  Formal approximations of the trait distributions moments in the regime of small variance $\varepsilon^2 \ll 1$

This appendix is dedicated to formally explain (12). We remove the time and the deme dependency for the sake of clarity. We denote $n_\varepsilon$ the trait distribution density, $N_\varepsilon$ the size of the population, $\overline{z}_\varepsilon$ the mean trait, $\sigma_\varepsilon^2$ the variance and $\psi_\varepsilon$ the third central moment. Let us also recall that the computations are performed using the exponential formalism introduced in (10) while considering the following formal expansion of $u_\varepsilon$ in the regime of small variance:

$$u_\varepsilon = u + \varepsilon^2 v + \mathcal{O}(\varepsilon^4).$$

**Size of population.** We have:

$$N_\varepsilon = \int_\mathbb{R} n_\varepsilon(z)\, dz$$

$$= \int_\mathbb{R} \frac{1}{\sqrt{2\pi}\varepsilon} e^{-\frac{(z-z^*)^2}{2\varepsilon^2}} e^{-u(z) - \varepsilon^2 v(z) + \mathcal{O}(\varepsilon^4)} dz$$

$$= \int_\mathbb{R} \frac{e^{-\frac{y^2}{2}}}{\sqrt{2\pi}} e^{-u(z^*+\varepsilon y) - \varepsilon^2 v(z^*+\varepsilon y) + \mathcal{O}(\varepsilon^4)} dy \qquad \left(y := \frac{z-z^*}{\varepsilon}\right)$$

$$= \int_\mathbb{R} \frac{e^{-\frac{y^2}{2}}}{\sqrt{2\pi}} e^{-[u(z^*) + \varepsilon y u'(z^*) + \frac{\varepsilon^2 y^2}{2} u''(z^*) + \frac{\varepsilon^3 y^3}{6} u'''(z^*) + \mathcal{O}(\varepsilon^4)] - \varepsilon^2 v(z^*) - \varepsilon^3 y v'(z^*) + \mathcal{O}(\varepsilon^4)} dy$$

$$= \int_\mathbb{R} \frac{e^{-\frac{y^2}{2}}}{\sqrt{2\pi}} e^{-u(z^*)} e^{-\left[\varepsilon y u'(z^*) + \varepsilon^2 \left[\frac{y^2 u''(z^*)}{2} + v(z^*)\right] + \varepsilon^3 \left[\frac{y^3}{6} u'''(z^*) + y v'\right] + \mathcal{O}(\varepsilon^4)\right]} dy$$

$$= \int_\mathbb{R} \frac{e^{-\frac{y^2}{2}}}{\sqrt{2\pi}} e^{-u(z^*)} \left[1 - \varepsilon y u'(z^*) - \varepsilon^2 \left[\frac{y^2 u''(z^*)}{2} + v(z^*)\right] - \varepsilon^3 \left[\frac{y^3}{6} u'''(z^*) - y v'(z^*)\right]\right.$$
$$\left. + \frac{1}{2}\left[\varepsilon^2 y^2 u'(z^*)^2 + \varepsilon^3 \left[y^3 u'(z^*) u''(z^*) + 2 y u'(z^*) v(z^*)\right]\right] - \frac{\varepsilon^3 y^3 u'(z^*)^3}{6} + \mathcal{O}(\varepsilon^4)\right]$$

$$= e^{-u(z^*)} \left[1 + \varepsilon^2 \left[\frac{u'^2(z^*)}{2} - \frac{u''(z^*)}{2} - v(z^*)\right]\right] + \mathcal{O}(\varepsilon^4),$$

from the computations of the moments of a Gaussian.

**Mean trait.** Similarly as above, we have:

$$\begin{aligned}
\overline{z}_\varepsilon &= \int_\mathbb{R} z \frac{n_\varepsilon}{N_\varepsilon} dz \\
&= \frac{1}{N_\varepsilon} \int_\mathbb{R} z \frac{1}{\sqrt{2\pi}\varepsilon} e^{-\frac{(z-z^*)^2}{2\varepsilon^2}} e^{-u(z)-\varepsilon^2 v(z) + \mathcal{O}(\varepsilon^4)} dz \\
&= \frac{1}{N_\varepsilon} \int_\mathbb{R} (z^* + \varepsilon y) \frac{e^{-\frac{y^2}{2}}}{\sqrt{2\pi}} e^{-u(z^*+\varepsilon y) - \varepsilon^2 v(z^*+\varepsilon y) + \mathcal{O}(\varepsilon^4)} dy, \qquad \left(y := \frac{z-z^*}{\varepsilon}\right) \\
&= \frac{1}{N_\varepsilon} \int_\mathbb{R} (z^* + \varepsilon y) \frac{e^{-\frac{y^2}{2}}}{\sqrt{2\pi}} e^{-u(z^*)} \left[ 1 - \varepsilon y u'(z^*) + \varepsilon^2 \left[ \frac{y^2 u'(z^*)^2}{2} - \frac{y^2 u''(z^*)}{2} - v(z^*) \right] \right. \\
&\qquad \left. + \varepsilon^3 \left[ -\frac{y^3}{6} u'''(z^*) - y v'(z^*) + \frac{y^3 u'(z^*) u''(z^*)}{2} + y u'(z^*) v(z^*) - \frac{3 y^3 u'(z^*)^3}{6} \right] + \mathcal{O}(\varepsilon^4) \right] \\
&= \frac{e^{-u(z^*)} \left[ z^* \left( 1 + \varepsilon^2 \left[ \frac{u'^2(z^*)}{2} - \frac{u''(z^*)}{2} - v(z^*) \right] \right) - \varepsilon^2 u'(z^*) \right] + \mathcal{O}(\varepsilon^4)}{e^{-u(z^*)} \left( 1 + \varepsilon^2 \left[ \frac{u'^2(z^*)}{2} - \frac{u''(z^*)}{2} - v(z^*) \right] \right) + \mathcal{O}(\varepsilon^4)} \\
&= z^* - \varepsilon^2 u'(z^*) + \mathcal{O}(\varepsilon^4).
\end{aligned}$$

**Variance.** Using the previous formal computations and methodology, we get:

$$\begin{aligned}
\sigma_\varepsilon^2 &= \frac{1}{N_\varepsilon} \int_\mathbb{R} (z - \overline{z}_\varepsilon)^2 n_\varepsilon(z) dz \\
&= \frac{1}{N_\varepsilon} \int_\mathbb{R} \left[ (z-z^*)^2 + (z^* - \overline{z}_\varepsilon)^2 + 2(z-z^*)(z^* - \overline{z}_\varepsilon) \right] n_\varepsilon(z) dz \\
&= \frac{1}{N_\varepsilon} \int_\mathbb{R} \left[ \varepsilon^2 y^2 + 2\varepsilon^3 y u'(z^*) + \mathcal{O}(\varepsilon^4) \right] \left[ 1 - \varepsilon y u' + \mathcal{O}(\varepsilon^2) \right] e^{-u(z^*)} \frac{e^{-\frac{y^2}{2}}}{\sqrt{2\pi}} dy \\
&= \frac{\varepsilon^2 e^{-u(z^*)}}{e^{-u(z^*)} \left[ 1 + \mathcal{O}(\varepsilon^2) \right]} \\
&= \varepsilon^2 + \mathcal{O}(\varepsilon^4).
\end{aligned}$$

**Third central moment.** We compute, using the same change in variable $y := \frac{z-z^*}{\varepsilon}$:

$$\begin{aligned}
\psi_\varepsilon^3 &= \frac{1}{N_\varepsilon} \int_\mathbb{R} (z - \overline{z}_\varepsilon)^3 n_\varepsilon(z) dz \\
&= \frac{1}{N_\varepsilon} \int_\mathbb{R} \left[ (z-z^*)^3 + (z^* - \overline{z}_\varepsilon)^3 + 3(z-z^*)^2(z^*-\overline{z}_\varepsilon) + 3(z-z^*)(z^*-\overline{z}_\varepsilon)^2 \right] n_\varepsilon(z) dz \\
&= \frac{1}{N_\varepsilon} \int_\mathbb{R} \left[ \varepsilon^3 y^3 + \mathcal{O}(\varepsilon^4) \right] \left[ e^{-u(z^*)} + \mathcal{O}(\varepsilon) \right] dz \\
&= \mathcal{O}(\varepsilon^4).
\end{aligned}$$

# E Fast/slow system: proof of Theorem 3.1

This appendix is dedicated to prove Theorem 3.1.

Let $(z_0^*, \overline{Y}_0^*) \in \mathbb{R} \times \Omega$ (we recall that $\Omega = (\mathbb{R}_+^*)^2 \times \mathbb{R}$) be on the slow manifold, ie. such that $G(z_0^*, \overline{Y}_0^*, 0) = 0$. From Lemma 6 of fast relaxation towards the slow manifold, the jacobian matrix $J_G(z_0^*, \overline{Y}_0^*)$ is invertible. Consequently, the implicit function theorem gives us $U$ open neighbourhood of $z_0^*$ in $\mathbb{R}$, $V$ open neighbourhood of $(z_0^*, \bar{Y}_0^*)$ in $\mathbb{R} \times \Omega$ and $\phi \in C^\infty(U, V)$ such that :
$$\forall (z^*, \bar{Y}^*) \in V, \, G(z^*, \bar{Y}^*, 0) = 0 \implies \bar{Y}^* = \phi(z^*).$$

Hence, we can define a notation that we shall use henceforth:
$$\forall z \in U, J_z := J_G(z, \phi(z)).$$

If $K$ is a compact subset of $U$ such that $z_0^* \in \mathring{K}$, we can define the Cauchy problem $(E_0)$ by the following :

$$(E_0) \quad \begin{cases} \frac{dz^*}{dt} = -2gz^*(t) + F(\phi(z^*(t))), \\ z^*(0) = z_0^*, \end{cases} \tag{28}$$

for $t \leq t^*$, that we define as the following:
$$t^* := \inf\{t > 0, z^*(t) \notin K\}.$$

It is similar to (20) with the initial conditions $(z^*(0), \bar{Y}_0^*) = (z_0^*, \phi(z_0^*))$. A essential part of the proof relies in the fact that we can define the following uniform positive constant, thanks to Lemma 6 of fast relaxation:

$$\lambda_K = -\frac{1}{2} \max_{z \in K} \{\lambda \in \mathbf{Sp}(J_z)\} > 0.$$

As the first step, we state the following lemma whose proof will be provided at the end of this appendix. It defines a uniform control constant $\gamma > 0$:

**Lemma 11.** *There exists $\gamma > 0$ such that:*
$$\max_{z \in K, s \geq 0} \left\vert\!\left\vert\!\left\vert e^{\lambda_K s} e^{J_z s} \right\vert\!\right\vert\!\right\vert \leq \gamma.$$

$(\left\vert\!\left\vert\!\left\vert \cdot \right\vert\!\right\vert\!\right\vert_{\mathcal{M}_3(\mathbb{R})}$ *is noted* $\left\vert\!\left\vert\!\left\vert \cdot \right\vert\!\right\vert\!\right\vert)$

The next step is to show the convergence of solutions of $(P_\varepsilon)$ (19) towards those of $(P_0)$ (20) on a time interval, yet to be defined, that will be shown to be uniform with regard to $\varepsilon$ and the initial conditions, provided that they are small enough. For that purpose, it is more convenient to consider the system $(R_\varepsilon)$ verified by the residuals $r_z^\varepsilon(t) = z_\varepsilon(t) - z^*(t)$ and $r_Y^\varepsilon(t) = \bar{Y}_\varepsilon(t) - \bar{Y}^*(t)$:

$$(R_\varepsilon) \quad \begin{cases} \varepsilon^2 \frac{dr_Y^\varepsilon}{dt} = G(z^*(t) + r_z^\varepsilon(t), \bar{Y}^*(t) + r_Y^\varepsilon(t), \varepsilon^2) - G(z^*(t), \bar{Y}^*(t), 0) - \varepsilon^2 \frac{d\bar{Y}^*}{dt} + \varepsilon^2 \nu_{N,\varepsilon}(t), \\ \frac{dr_z^\varepsilon}{dt} = -2gr_z^\varepsilon(t) + F(\bar{Y}^*(t) + r_Y^\varepsilon) - F(\bar{Y}^*(t)) + \varepsilon^2 \nu_{z,\varepsilon}(t), \\ (r_z^\varepsilon(0), r_Y^\varepsilon(0)) = (z_0^\varepsilon - z_0^*, \bar{Y}_0^\varepsilon - \bar{Y}_0^*), \end{cases} \tag{29}$$

and introduce some further definitions.

Because $K$ is a compact set, there exists $\delta_K > 0$ such that the following set is a compact subset of $V$:
$$\bar{K}_{\delta_K} = \{(z, \bar{Y}) \in \mathbb{R} \times \Omega | \exists z^* \in K, |(z, \bar{Y}) - (z^*, \phi(z^*))| \leq \delta_K\} \subset V.$$

Let us consider from now $(z_0^\varepsilon, N_0^\varepsilon) \in \bar{K}_{\delta_K}$. Then we define $\Delta = \min\left(\frac{\lambda_K}{4C\gamma}, \delta_K\right)$ and $T = \min(t^*, \frac{\lambda_K}{4C'\gamma})$, where:

$$C = \max\left(\|\partial_{\bar{Y}}^2 G\|_{\infty, \bar{K}_{\delta_K} \times [0,1]}, \|\partial_z G\|_{\infty, \bar{K}_{\delta_K} \times [0,1]}, \|\partial_\varepsilon G\|_{\infty, \bar{K}_{\delta_K} \times [0,1]}, \|\partial_{\bar{Y}} F\|_{\infty, \Pi_\Omega(\bar{K}_{\delta_K})}\right)$$

and :

$$C' = \max_{t \leq t^*}\left\|\partial_t J_{z^*(t)}\right\|,$$

where $\Pi_\Omega$ is the projection from $\mathbb{R} \times \Omega$ on $\Omega$. One can notice from these definitions and from Lemma 11, that $\gamma, \Delta, T, \lambda_K, C, C'$ do not depend on $\varepsilon$ and are uniform on $[0, t^*]$. Specifically taking $\Delta \leq \frac{\lambda_K}{4C\gamma}$ and $T \leq \frac{\lambda_K}{4C'\gamma}$ will turn out to be important in the proof.

On the time region $[0, T]$, we will show that we can control explicitly the various perturbed terms that arise. We can now state the following proposition, whose proof constitutes the core of the resolution of the problem:

**Proposition E.1.** As $\max(\varepsilon, |r_z^\varepsilon(0)|, |r_Y^\varepsilon(0)|) \to 0$, $(\bar{Y}_\varepsilon, z_\varepsilon)$ converges toward $(\bar{Y}^*, z^*)$ uniformly on $[0, T]$.

For the final step, we will show that we can reiterate the process on each interval of time $[jT, \min\{(j+1)T, t^*\}]$ with $\forall j \leq \lfloor \frac{t^*}{T} \rfloor, jT \leq t_\varepsilon^*$. Thus, for sufficiently small $\varepsilon$ and initial conditions, the control remains valid until $t^*$, hence Theorem 1.

For convenience, we will denote by $f * g(t)$ the convolution product of $f$ and $g$ at time $t > 0$:

$$f * g(t) = \int_0^t f(\tau) g(t-\tau) d\tau.$$

*Proof of Proposition E.1.* Let $\varepsilon \in ]0,1]$. Let us define an auxiliary time $t_\varepsilon^*$:

$$t_\varepsilon^* = \min\left(t^*, \inf\{t > 0, |r_z^\varepsilon| + |r_Y^\varepsilon| > \Delta\}\right).$$

It ensures that the perturbed trajectory stays inside of $\bar{K}_{\delta_K}$ when $t \leq t_\varepsilon^*$.

Let us highlight the main steps of the proof:

1. preliminary controls on $r_Y^\varepsilon$ by $|r_Y^\varepsilon(0)|$ and $\frac{1}{\varepsilon^2}|r_z^\varepsilon| * e^{-\frac{\lambda_K}{2\varepsilon^2}\cdot}$ thanks to the regularity of $G$, the fast relaxation properties (Lemma 6 and Lemma 11) and Gronwall's lemma.

2. control $|r_z^\varepsilon|$ by $|r_z^\varepsilon(0)|$ and $|r_Y^\varepsilon|$.

3. finish the control on $r_Y^\varepsilon$ by using the latter and Gronwall's lemma.

**1.** For $t \leq \min(T, t_\varepsilon^*)$, we can introduce new terms in the equation from (29) on $r_Y^\varepsilon$:

$$\begin{aligned}\frac{dr_Y^\varepsilon}{dt} &= \frac{J_{z^*(0)}}{\varepsilon^2} r_Y^\varepsilon + \frac{1}{\varepsilon^2}\left[G(z^*(t), \bar{Y}^*(t) + r_Y^\varepsilon(t), 0) - G(z^*(t), \bar{Y}^*(t), 0) - J_{z^*(0)} r_Y^\varepsilon\right] \\ &\quad + \frac{1}{\varepsilon^2}\left[G(z^*(t) + r_z^\varepsilon(t), \bar{Y}^*(t) + r_Y^\varepsilon(t), \varepsilon^2) - G(z^*(t), \bar{Y}^*(t) + r_Y^\varepsilon(t), 0)\right] \\ &\quad - \phi'(z^*(t))(-2gz^*(t) + F(\phi(z^*(t)))) + \nu_{N,\varepsilon}(t) \\ &= \frac{J_{z^*(0)}}{\varepsilon^2} r_Y^\varepsilon + A_1(t) + A_2(t) + A_3(t).\end{aligned}$$

Since $t \leq \min(T, t_\varepsilon^*)$ and $G$ is $C^\infty$ on $\bar{K}_{\delta_K} \times [0, 1]$, we can control $A_1$:

$$|A_1(t)| \leq \frac{1}{\varepsilon^2}\left[|G(z^*(t), \bar{Y}^*(t) + r_Y^\varepsilon(t), 0) - G(z^*(t), \bar{Y}^*(t), 0) - J_{z^*(t)} r_Y^\varepsilon|\right]$$
$$+ \frac{1}{\varepsilon^2}\left[\left\|\!\left|J_{z^*(t)} - J_{z^*(0)}\right|\!\right\| |r_Y^\varepsilon(t)|\right]$$

$$\leq \frac{1}{\varepsilon^2}\left[\|\partial_Y^2 G\|_{\infty, \bar{K}_{\delta_K} \times [0,1]} |r_Y^\varepsilon(t)|^2 + T \max_{t \leq t^*}\left\|\!\left|\partial_t J_{z^*(t)}\right|\!\right\| |r_Y^\varepsilon(t)|\right]$$

$$\leq \frac{1}{\varepsilon^2}(C\Delta + C'T)|r_Y^\varepsilon(t)|,$$

and $A_2$:

$$|A_2(t)| = \frac{1}{\varepsilon^2}|G(z^*(t) + r_z^\varepsilon(t), \bar{Y}^*(t) + r_Y^\varepsilon(t), \varepsilon^2) - G(z^*(t), \bar{Y}^*(t) + r_Y^\varepsilon(t), 0)|$$

$$\leq \frac{1}{\varepsilon^2}\left[\|\partial_z G\|_{\infty, \bar{K}_{\delta_K} \times [0,1]} |r_z^\varepsilon(t)| + \varepsilon^2 \|\partial_\varepsilon G\|_{\infty, \bar{K}_{\delta_K} \times [0,1]}\right] \leq \frac{C}{\varepsilon^2}(|r_z^\varepsilon(t)| + \varepsilon^2),$$

and $A_3$:

$$|A_3(t)| = |-\phi'(z^*(t))(-2gz^*(t) + F(\phi(z^*(t)))) + \nu_{N,\varepsilon}(t)| \leq C'',$$

for some constant $C''$ independent of $\varepsilon$ and $z^*(0) \in K$. Using Duhamel formulas, we get, for $t \leq \min(T, t_\varepsilon^*)$:

$$r_Y^\varepsilon(t) = e^{\frac{J_{z^*(0)} t}{\varepsilon^2}} r_Y^\varepsilon(0) + \left[e^{\frac{J_{z^*(0)} \cdot}{\varepsilon^2}} * (A_1 + A_2 + A_3)\right](t). \tag{30}$$

Hence, applying Lemma 11 yields:

$$|r_Y^\varepsilon(t)| \leq \gamma |r_Y^\varepsilon(0)| e^{-\frac{\lambda_K t}{\varepsilon^2}} + \frac{\gamma}{\varepsilon^2}\left[(C|r_z^\varepsilon| + (C\Delta + C'T)|r_Y^\varepsilon|) * e^{-\frac{\lambda_K \cdot}{\varepsilon^2}}\right](t) + \varepsilon^2 \gamma \frac{C + C''}{\lambda_K}$$

$$\leq A^{r_z^\varepsilon}(t) + \frac{\gamma(C\Delta + C'T)}{\varepsilon^2} \int_0^t |r_Y^\varepsilon(\tau)| e^{\frac{\lambda_K}{\varepsilon^2}(\tau - t)} d\tau,$$

where $A^{r_z^\varepsilon}(t) := \gamma|r_Y^\varepsilon(0)|e^{-\frac{\lambda_K t}{\varepsilon^2}} + \frac{\gamma C}{\varepsilon^2}\left(|r_z^\varepsilon| * e^{-\frac{\lambda_K \cdot}{\varepsilon^2}}\right)(t) + \varepsilon^2 \gamma \frac{C+C''}{\lambda_K}$.

Applying Gronwall inequality to $r_Y^\varepsilon(t) e^{\frac{\lambda_K t}{\varepsilon^2}}$ yields:

$$|r_Y^\varepsilon(t)| \leq A^{r_z^\varepsilon}(t) + \frac{\gamma(C\Delta + C'T)}{\varepsilon^2}\left[A^{r_z^\varepsilon} * e^{\left(\frac{-\lambda_K}{\varepsilon^2} + \frac{\gamma(C\Delta + C'T)}{\varepsilon^2}\right) \cdot}\right](t). \tag{31}$$

Having fixed $\Delta \leq \frac{\lambda_K}{4C\gamma}$ and $T \leq \frac{\lambda_K}{4C'\gamma}$ in the preliminaries ensures that $e^{\left(\frac{-\lambda_K}{\varepsilon^2} + \frac{\gamma(C\Delta + C'T)}{\varepsilon^2}\right) \cdot}$ defines a negative exponential term, that we can dominate by $e^{-\frac{\lambda_K}{2\varepsilon^2} \cdot}$. Hence:

$$|r_Y^\varepsilon(t)| \leq A^{r_z^\varepsilon}(t) + \left[A^{r_z^\varepsilon} * \frac{\lambda_K}{2\varepsilon^2} e^{-\frac{\lambda_K}{2\varepsilon^2} \cdot}\right](t). \tag{32}$$

Making $A^{r_z^\varepsilon}$ explicit gives:

$$|r_Y^\varepsilon(t)| \leq \gamma |r_Y^\varepsilon(0)|e^{-\frac{\lambda_K t}{\varepsilon^2}} + \frac{\gamma C}{\varepsilon^2}|r_z^\varepsilon| * e^{-\frac{\lambda_K}{\varepsilon^2}\cdot}(t) + \varepsilon^2 \gamma \frac{C+C''}{\lambda_K}$$
$$+ \left[\left(\gamma|r_Y^\varepsilon(0)|e^{-\frac{\lambda_K}{\varepsilon^2}\cdot} + \frac{\gamma C}{\varepsilon^2}\left[|r_z^\varepsilon| * e^{-\frac{\lambda_K}{\varepsilon^2}\cdot}\right] + \varepsilon^2\gamma\frac{C+C''}{\lambda_K}\right) * \left(\frac{\lambda_K}{2\varepsilon^2}e^{-\frac{\lambda_K}{2\varepsilon^2}\cdot}\right)\right](t)$$

$$\leq \gamma|r_Y^\varepsilon(0)|\left[e^{-\frac{\lambda_K t}{\varepsilon^2}} + e^{-\frac{\lambda_K}{\varepsilon^2}\cdot} * \left(\frac{\lambda_K}{2\varepsilon^2}e^{-\frac{\lambda_K}{2\varepsilon^2}\cdot}\right)(t)\right] + \varepsilon^2\gamma\frac{C+C''}{\lambda_K}\left(\left(1 + \int_0^t \frac{\lambda_K}{2\varepsilon^2}e^{-\frac{\lambda_K}{2\varepsilon^2}(\tau-t)}dt\right)\right.$$
$$+ \frac{\gamma C}{\varepsilon^2}|r_z^\varepsilon| * \left(e^{-\frac{\lambda_K}{\varepsilon^2}\cdot} + e^{-\frac{\lambda_K}{\varepsilon^2}\cdot} * \frac{\lambda_K}{2\varepsilon^2}e^{-\frac{\lambda_K}{2\varepsilon^2}\cdot}\right)(t), \quad (33)$$

thanks to the associativity of the convolution product. One can compute that, for $t \geq 0$:

$$e^{-\frac{\lambda_K t}{\varepsilon^2}} + e^{-\frac{\lambda_K}{\varepsilon^2}\cdot} * \left(\frac{\lambda_K}{2\varepsilon^2}e^{-\frac{\lambda_K}{2\varepsilon^2}\cdot}\right)(t) = e^{-\frac{\lambda_K t}{\varepsilon^2}} + \frac{\lambda_K}{2\varepsilon^2}\int_0^t e^{-\frac{\lambda_K}{\varepsilon^2}\tau}e^{-\frac{\lambda_K}{2\varepsilon^2}(t-\tau)}d\tau$$
$$= e^{-\frac{\lambda_K t}{\varepsilon^2}} + \frac{\lambda_K}{2\varepsilon^2}\int_0^t e^{-\frac{\lambda_K}{2\varepsilon^2}(t+\tau)}d\tau = e^{-\frac{\lambda_K}{2\varepsilon^2}t}.$$

Hence, replacing those terms in (33) yields:

$$|r_Y^\varepsilon(t)| \leq \gamma|r_Y^\varepsilon(0)|e^{-\frac{\lambda_K}{2\varepsilon^2}t} + 2\varepsilon^2\gamma\frac{C+C''}{\lambda_K} + \frac{C\gamma}{\varepsilon^2}|r_z^\varepsilon| * e^{-\frac{\lambda_K}{2\varepsilon^2}\cdot}(t). \quad (34)$$

**2.** The next step is to gain similarly some control on $|r_z^\varepsilon|$. Using Duhamel formula on the equation from (29) on $r_z^\varepsilon$ gives, for $t \leq \min(T, t_\varepsilon^*)$:

$$r_z^\varepsilon(t) = r_z^\varepsilon(0)e^{-2gt} + \left(\left[F(N^* + r_Y^\varepsilon) - F(N^*) + \varepsilon^2 \nu_{z,\varepsilon}\right] * e^{-2g\cdot}\right)(t),$$

which yields:

$$|r_z^\varepsilon(t)| \leq |r_z^\varepsilon(0)|e^{-2gt} + \varepsilon^2\frac{\|\nu_{z,\varepsilon}\|_\infty}{2g} + \|\partial_{\bar{Y}} F\|_{\infty, \Pi_\Omega(\bar{K}_{\delta_K})}\left(|r_Y^\varepsilon| * e^{-2g\cdot}\right)(t).$$

Hence:

$$|r_z^\varepsilon(t)| \leq |r_z^\varepsilon(0)|e^{-2gt} + \varepsilon^2\frac{\|\nu_{z,\varepsilon}\|_\infty}{2g} + C\left(|r_Y^\varepsilon| * e^{-2g\cdot}\right)(t). \quad (35)$$

At that point, it is clear that it is sufficient to control $|r_Y^\varepsilon|$ and $|r_z^\varepsilon(0)|$ in order to control $|r_z^\varepsilon(t)|$ for sufficiently small $\varepsilon$.

**3.** Plugging the latter in (34) gives:

$$|r_Y^\varepsilon(t)| \leq \gamma|r_Y^\varepsilon(0)|e^{-\frac{\lambda_K}{2\varepsilon^2}t} + \frac{C\gamma}{\varepsilon^2}|r_z^\varepsilon(0)|\left(e^{-2g\cdot} * e^{-\frac{\lambda_K}{2\varepsilon^2}\cdot}\right)(t) + \varepsilon^2\frac{C\gamma\|\nu_{z,\varepsilon}\|_\infty}{\lambda_K g}$$
$$+ 2\varepsilon^2\gamma\frac{C+C''}{\lambda_K} + \frac{\gamma C^2}{\varepsilon^2}\left[|r_Y^\varepsilon| * \left(e^{-2g\cdot} * e^{-\frac{\lambda_K}{2\varepsilon^2}\cdot}\right)\right](t). \quad (36)$$

Similarly as the computation above, we have, for $\varepsilon^2 < \min(\frac{\lambda_K}{8g}, 1)$ and $t \geq 0$:

$$e^{-2g\cdot} * e^{-\frac{\lambda_K}{2\varepsilon^2}\cdot}(t) = \frac{1}{\frac{\lambda_K}{2\varepsilon^2} - 2g}\left(e^{-2gt} - e^{-\frac{\lambda_K}{2\varepsilon^2}t}\right) \leq \frac{4\varepsilon^2}{\lambda_K}e^{-2gt}.$$

Hence, for $\varepsilon^2 < \min(\frac{\lambda_K}{8g}, 1)$, we get from (36):

$$|r_Y^\varepsilon(t)| \leq \gamma |r_Y^\varepsilon(0)|e^{-\frac{\lambda_K}{2\varepsilon^2}t} + \frac{2\gamma C}{\lambda_K}|r_z^\varepsilon(0)|e^{-2gt} + \varepsilon^2 \frac{C\gamma \|\nu_{z,\varepsilon}\|_\infty}{\lambda_K g} + 2\varepsilon^2 \gamma \frac{C+C''}{\lambda_K}$$
$$+ \frac{2\gamma C^2}{\lambda_K}\left(|r_Y^\varepsilon| * e^{-2g\cdot}\right)(t)$$
$$\leq C_0^\varepsilon(t) + \frac{2\gamma C^2}{\lambda_K}\left(|r_Y^\varepsilon| * e^{-2g\cdot}\right)(t),$$

where we define: $C_0^\varepsilon(t) := \gamma|r_Y^\varepsilon(0)|e^{-\frac{\lambda_K}{2\varepsilon^2}t} + \frac{2\gamma C}{\lambda_K}|r_z^\varepsilon(0)|e^{-2gt} + \varepsilon^2\frac{C\gamma\|\nu_{z,\varepsilon}\|_\infty}{\lambda_K g} + 2\varepsilon^2\gamma\frac{C+C''}{\lambda_K}$.

Using once again Gronwall inequality on $|r_Y^\varepsilon|e^{2g\cdot}$ yields:

$$|r_Y^\varepsilon(t)| \leq C_0^\varepsilon(t) + \frac{2\gamma C^2}{\lambda_K}\left(C_0^\varepsilon * e^{\left(-2g+\frac{2\gamma C^2}{\lambda_K}\right)\cdot}\right)(t). \tag{37}$$

Recalling that:

$$C_0^\varepsilon(t) = \gamma|r_Y^\varepsilon(0)|e^{-\frac{\lambda_K}{2\varepsilon^2}t} + \frac{2\gamma C}{\lambda_K}|r_z^\varepsilon(0)|e^{-2gt} + \varepsilon^2\frac{C\gamma\|\nu_{z,\varepsilon}\|_\infty}{\lambda_K g} + 2\varepsilon^2\gamma\frac{C+C''}{\lambda_K},$$

we get that, thanks to (37) and (35), for a given $0 < \delta < \Delta$, there exists $\eta_\delta > 0$ depending only on $\delta, g, m, K, t^*, F, G, \|\nu_{z,\varepsilon}\|_\infty$ such that:

$$\forall(\varepsilon, |r_Y^\varepsilon(0)|, |r_z^\varepsilon(0)|) \in [0, \eta_\delta]^3, \max_{t \leq \min(T, t_\varepsilon^*)} |r_Y^\varepsilon(t)| + |r_z^\varepsilon(t)| \leq \delta.$$

Recalling that $t_\varepsilon^* = \min\left(t^*, \inf\{t > 0, |r_z^\varepsilon| + |r_Y^\varepsilon| > \Delta\}\right)$, we get that $T \leq t_\varepsilon^*$, for $\delta < \Delta$ and $(\varepsilon, |r_Y^\varepsilon(0)|, |r_z^\varepsilon(0)|) \in [0, \eta_\delta]^3$. Consequently, the convergence is uniform on $[0, T]$. □

*Proof of Theorem 1.* One can notice that the control obtained in the proof of Proposition 1 can be applied on any time interval $[a, a+T]$ with $a \in [0, t^* - T]$, provided that $(\varepsilon, |r_Y^\varepsilon(a)|, |r_z^\varepsilon(a)|)$ are small enough. Therefore, we can reiterate the control a finite number of times on the intervals $[jT, \min\{(j+1)T, t^*\}]$ with $\forall j \leq \lfloor \frac{t^*}{T} \rfloor$. Hence, the uniform convergence on $[0, t^*]$. □

*Proof of Lemma 11.* Recall that for all $z \in K$, $J_z$ has real negative eigenvalues, uniformly bounded over $K$ by $-2\lambda_K < -\lambda_K$. Let us define, for $z \in K$:

$$f_{\lambda_K, z} : \mathbb{R}_+ \to \mathbb{R}_+, \quad s \mapsto \left\|e^{J_z s}e^{\lambda_K s}\right\|.$$

For all $z \in \mathbb{K}$, $f_{\lambda_K, z}$ is continuous. Moreover, Theorem 2.34 of Chicone 1999 ensures that $f_{l,z}$ is bounded for all $l < 2\lambda_K$.

We can thus define:

$$\Gamma_{\lambda_K} : K \to \mathbb{R}_+^*, \quad z \mapsto \max_{s \geq 0} f_{\lambda_K, z}(s).$$

Let us show that $\Gamma_{\lambda_K}$ is a continuous function. Let $z_0 \in K$ and $\varepsilon > 0$.

One can first notice that, for $s \geq 0$:

$$f_{\lambda_K, z}(s) = f_{\frac{3\lambda_K}{2}, z}(s)e^{-\frac{\lambda_K}{2}s} < \Gamma_{\frac{3\lambda_K}{2}, z}e^{-\frac{\lambda_K}{2}s}.$$

Thus, $f_{\lambda_K, z}$ vanishes when $s$ goes to infinity. As a consequence, there exists $s_0 \geq 0$ such that:

$$\Gamma_{\lambda_K}(z_0) = \left\|e^{J_{z_0} s_0}e^{\lambda_K s_0}\right\|.$$

Furthermore, for $l \in ]\lambda_K, 2\lambda_K[$, we have:
$$\Gamma_l(z_0) = \left|\!\left|\!\left| e^{J_{z_0} s_0} e^{l s_0} \right|\!\right|\!\right| = \Gamma_{\lambda_K}(z_0) e^{(l-\lambda_K) s_0}.$$

We can therefore choose $l \in ]\lambda_K, 2\lambda_K[$ such that $\Gamma_{\lambda_K}(z_0) \leq \Gamma_l(z_0) \leq \Gamma_{\lambda_K}(z_0) + \varepsilon$.

As $z \mapsto J_z$ is a continuous function, there exists $\delta > 0$ that ensures that for if $z \in K$ and $|z - z_0| \leq \delta$, then:
$$\left|\!\left|\!\left| J_z - J_{z_0} \right|\!\right|\!\right| < \frac{l - \lambda_K}{2\Gamma_l(z_0)}.$$

Let us consider such a $z$.

As $e^{J_z s}$ is solution of the ODE : $y' = J_{z_0} y + (J_z - J_{z_0})y$, we obtain, for $s \geq 0$:
$$e^{J_z s} = e^{J_{z_0} s} + e^{J_{z_0} \cdot} * (J_z - J_{z_0}) e^{J_z \cdot}(s).$$

Hence :
$$\left|\!\left|\!\left| e^{J_z t} \right|\!\right|\!\right| \leq \Gamma_l(z_0) e^{-ls} + \frac{l - \lambda_K}{2} \left|\!\left|\!\left| e^{J_z \cdot} \right|\!\right|\!\right| * e^{-l \cdot}$$

From applying Gronwall's inequality to $t \mapsto \left|\!\left|\!\left| e^{J_z s} \right|\!\right|\!\right| e^{ls}$, it comes that, for $s \geq 0$:
$$\left|\!\left|\!\left| e^{J_z s} \right|\!\right|\!\right| \leq \Gamma_l(z_0) e^{-\left(l - \frac{l-\lambda_K}{2}\right)t} \leq \Gamma_l(z_0) e^{-\left(\frac{l+\lambda_K}{2}\right)s}$$
$$\leq [\Gamma_{\lambda_K}(z_0) + \varepsilon] e^{-\lambda_K s}.$$

Hence:
$$\Gamma_{\lambda_K}(z) \leq \Gamma_{\lambda_K}(z_0) + \varepsilon.$$

Moreover, recall that $t_0$ was defined so that :
$$\Gamma_{\lambda_K}(z_0) = \left|\!\left|\!\left| e^{J_{z_0} s_0} e^{\lambda_K s_0} \right|\!\right|\!\right|.$$

Then, by continuity of $z \mapsto e^{J_z s_0}$, there exists $\delta' > 0$ that ensures that for $|z - z_0| \leq \delta'$, we have:
$$\left|\!\left|\!\left| e^{J_z s_0} e^{\lambda_K s_0} \right|\!\right|\!\right| \geq \left|\!\left|\!\left| e^{J_{z_0} s_0} e^{\lambda_K s_0} \right|\!\right|\!\right| - \varepsilon.$$

Hence:
$$\Gamma_{\lambda_K}(z) \geq \Gamma_{\lambda_K}(z_0) - \varepsilon.$$

In conclusion, if $|z - z_0| \leq \min(\delta, \delta')$, then $|\Gamma_{\lambda_K}(z) - \Gamma_{\lambda_K}(z_0)| \leq \varepsilon$. Hence $\Gamma_{\lambda_K}$ is continuous over $K$. Furthermore, as $K$ is a compact set, $\Gamma_{\lambda_K}$ is bounded, by $\gamma$. $\square$

# F  Proof of Proposition 3.1

This appendix is dedicated to the proof of Proposition 3.1.

*Proof.* Let $(g, m, z^*) \in \mathbb{R}_+^* \times \mathbb{R}_+^* \times \mathbb{R}_+$ be such that $P_{z^*}$ has a single positive root. From Lemma 1, this root defines a fast equilibrium if it is greater than $f_1(z^*)$. From Lemma 2, that is the case if and only if $f_1(z^*)$ is negative or $P_{z^*}(f_1(z^*))$ is negative.

First, regarding the sign of $f_1(z^*)$, we have:
$$f_1(z^*) < 0 \iff (z^* + 1)^2 < \frac{1 - m}{g},$$

which requires that $m < 1$. If $m < 1$ then:

$$f_1(z^*) < 0 \iff 0 \leq z^* < \sqrt{\frac{1-m}{g}} - 1,$$

which requires that $m + g < 1$. Hence:

$$f_1(z^*) < 0 \iff [m + g < 1] \wedge [z^* < \sqrt{\frac{1-m}{g}} - 1].$$

Next, regarding the sign of $P_{z^*}(f_1(z^*))$, we compute:

$$\begin{aligned} P_{z^*}(f_1(z^*)) &= f_1(z^*)f_2(z^*) - 1 \\ &= \left(1 + \frac{g}{m}(z^* + 1)^2 - \frac{1}{m}\right)\left(1 + \frac{g}{m}(z^* - 1)^2 - \frac{1}{m}\right) - 1 \\ &= \frac{g^2}{m^2}\left[z^{*4} + z^{*2}\frac{2(m-g-1)}{g} + \frac{(g-1)(2m+g-1)}{g^2}\right] \end{aligned}$$

Let us define:

$$Q(X) = X^2 + X\frac{2(m-g-1)}{g} + \frac{(g-1)(2m+g-1)}{g^2},$$

$z_1, z_2$ its two roots and $\Delta = \frac{4}{g^2}\left[m^2 - 4g(m-1)\right]$ its discriminant. From the computation above,

$$P_{z^*}(f_1(z^*)) < 0 \iff [\Delta > 0] \wedge [z^{*2} \in ]z_1, z_2[].$$

We have:

$$\begin{aligned} \Delta > 0 = &\iff m^2 - 4gm + 4g > 0 \\ &\iff [g < 1] \vee \left[[g \geq 1] \wedge \left[\left[0 < m < 2g\left(1 - \sqrt{1 - \frac{1}{g}}\right)\right] \vee \left[m > 2g\left(1 + \sqrt{1 - \frac{1}{g}}\right)\right]\right]\right] \end{aligned}$$

and:

$$z_1 z_2 = \frac{(g-1)(2m+g-1)}{g^2}, \quad z_1 + z_2 = \frac{2(g+1-m)}{g}.$$

Consequently:

⋄ if $g \geq 1$, then $2m + g - 1 > 0$ and then $z_1 z_2 \geq 0$. If, additionally, $m < 2g\left(1 - \sqrt{1 - \frac{1}{g}}\right)$, then $m < 2 \leq g + 1$ ($g \mapsto 2g - 2\sqrt{g^2 - g}$ is decreasing on $[1, +\infty[)$. Therefore, we get: $z_1 + z_2 > 0$ and thus, $z_2 > 0$ and $z_1 \geq 0$. At last, if $m > 2g\left(1 + \sqrt{1 - \frac{1}{g}}\right)$, then $m > 2g \geq g + 1$, which implies $z_1 + z_2 < 0$ and thus $z_1 < 0, z_2 \leq 0$.

⋄ if $g < 1$, then $z_1 + z_2 \geq 0$ if and only if $m \leq g + 1$ and $z_1 z_2 \geq 0$ if and only if $m \leq \frac{1-g}{2}$ (which is lower than $g + 1$).

Hence the result. □

# G  Proof of Lemma 9

This section is dedicated to proving Lemma 9, which concludes the proof of Proposition 4.2.

*Proof of Lemma 9.* Let $(m, g) \in \mathbb{R}_+^{*\,2}$ verify (25). Then, from the first part of the proof of Proposition 4.2, there exists a unique $\rho^* > 0$ that is solution of the equation in (23). Let us define $N_1^*$ and $N_2^*$ such as in (26). Then we have: $0 < \rho^* = \frac{N_2^*}{N_1^*}$. Thus:

$$N_1^* > 0 \iff N_2^* > 0 \iff \frac{1}{m}(N_1^* + N_2^*) > 0.$$

Borrowing once again the notations: $a = \frac{4g}{m}$, $b = \frac{1}{m}$ and $y^* = \rho^* + \frac{1}{\rho^*}$ (unique root of $S$ larger than 2), (26) leads to:

$$\frac{1}{m}(N_1^* + N_2^*) = 2\left(\frac{1}{m} - 1\right) + y^* - \frac{4g}{m}\frac{y^{*2} - 2}{y^{*2}}$$

$$= \frac{1}{y^{*2}}\left[y^{*3} + \left[\frac{1-2m}{m} + \frac{1}{m} - \frac{4g}{m}\right]y^{*2} + 2 \times \frac{4g}{m}\right]$$

$$= \frac{1}{y^{*2}}\left[S(y^*) + (\frac{1-2m}{m})y^{*2} + \frac{4g}{m}y^* + \frac{4g}{m}\right].$$

As $S(y^*) = 0$, we get:

$$N_1^* > 0 \iff N_2^* > 0 \iff (1-2m)y^{*2} + 4gy^* + 4g > 0.$$

This is always true whenever $m \leq \frac{1}{2}$. Otherwise, let us suppose henceforth that $2m > 1$. The condition above is equivalent to:

$$y^* < c + \sqrt{c^2 + 2c}, \quad \text{where: } c = \frac{2g}{2m-1} > 0.$$

Let us show that: $c + \sqrt{c^2 + 2c} \geq 2$. It is sufficient to show that: $c \geq \frac{2}{3}$, which is equivalent to having: $3g + 1 \geq 2m$. In this proof, we are considering $(m, g) \in \mathbb{R}_+^{*\,2}$ such that $1 + 2m < 5g$ and $4g(m-1) < m^2$. Let us show that such pairs verify $3g + 1 \geq 2m$:

- ⋄ if $g \leq 1$, then $m < \frac{5g-1}{2} \leq \frac{3g+1}{2}$.
- ⋄ if $g \geq 1$, then $m < 2g - 2\sqrt{g^2 - g}$ which is a decreasing function on $[1, +\infty[$, which takes the value 2 when $g = 1$. Hence it is always dominated by $g \mapsto \frac{3g+1}{2}$ on this interval.

Hence $c + \sqrt{c^2 + 2c} \geq \frac{2}{3} + \sqrt{\frac{4}{9} + \frac{4}{3}} = 2$. Therefore, as $y^*$ is the only root of $S$ greater than 2, we get the following equivalence:

$$y^* < c + \sqrt{c^2 + c} \iff S\left(c + \sqrt{c^2 + 2c}\right) > 0.$$

The rest of the proof is dedicated to examine the conditions on $(m, g)$ under which:

$$S\left(c + \sqrt{c^2 + 2c}\right) > 0.$$

Let us set $Q := \sqrt{c^2 + 2c} = \sqrt{4g\frac{g+2m-1}{(2m-1)^2}}$. Tedious computations done with the help of Mathematica show that: $S(c) = Q^2\left[\frac{g(4-6m)+(2m-1)^2}{m(2m-1)}\right]$, and we next compute:

$$S(c + Q) = S(c) + Q^2\left[3c + \frac{1-4g}{m}\right] + Q\left[Q^2 + 3c^2 + 2c\frac{(1-4g)}{m} - \frac{4g}{m}\right]$$

$$= Q^2 \left[ \frac{g(4-6m) + (2m-1)^2}{m(2m-1)} + \frac{6g}{2m-1} + \frac{1-4g}{m} \right] + Q \left[ 4c^2 + 2c\frac{(m+1-4g)}{m} - \frac{4g}{m} \right]$$

$$= Q \left[ 2Q \frac{(2m^2 - m - 4g(m-1))}{m(2m-1)} - \frac{4g(4g(m-1) + 2m^2 - 5m + 2)}{m(2m-1)^2} \right].$$

Hence:
$$S(c+Q) > 0$$
$$\iff Q\left(2m^2 - m - 4g(m-1)\right) > 2g\frac{(4g(m-1) + 2m^2 - 5m + 2)}{(2m-1)}$$
$$\iff \sqrt{g + 2m - 1}\left(2m^2 - m - 4g(m-1)\right) > 2\sqrt{g}\left(4g(m-1) + 2m^2 - 5m + 2\right).$$

Let us study different cases corresponding to different ranges of value of $m > \frac{1}{2}$.

If $\underline{m = 1}$, then the last line is equivalent to:
$$\sqrt{1+g} > -2\sqrt{g},$$
which is true for all $g > 0$.

If $\underline{\frac{1}{2} < m < 1}$, then:
$$4g(m-1) + 2m^2 - 5m + 2 = 4g(m-1) + 2(m-2)\left(m - \frac{1}{2}\right) < 0,$$
and:
$$2m^2 - m - 4g(m-1) = 2m\left(m - \frac{1}{2}\right) + 4g(1-m) > 0.$$

Hence, for all $g$ such that $1 + 2m < 5g$ and $m^2 > 4g(m-1)$:
$$\sqrt{g + 2m - 1}\left(2m^2 - m - 4g(m-1)\right) > 2\sqrt{g}\left(4g(m-1) + 2m^2 - 5m + 2\right).$$

If $\underline{m \geq 1}$, then:
$$2m^2 - m > m^2 > 4g(m-1).$$

Hence, if: $4g(m-1) + 2m^2 - 5m + 2 < 0$, then, for all $g$ such that $1 + 2m < 5g$ and $m^2 > 4g(m-1)$:
$$\sqrt{g + 2m - 1}\left(2m^2 - m - 4g(m-1)\right) > 2\sqrt{g}\left(4g(m-1) + 2m^2 - 5m + 2\right).$$

Otherwise, if $4g(m-1) + 2m^2 - 5m + 2 \geq 0$, then:
$$S(c+Q) > 0$$
$$\iff \sqrt{g + 2m - 1}\left(2m^2 - m - 4g(m-1)\right) > 2\sqrt{g}\left(4g(m-1) + 2m^2 - 5m + 2\right)$$
$$\iff \left(1 + \frac{2m-1}{g}\right)\left(2m^2 - 2 - 4g(m-1)\right)^2 > 4\left(4g(m-1) + 2m^2 - 5m + 2\right)^2.$$

Let us note $x := \frac{2m-1}{g}$. Then, the latter is equivalent to:

$$(1+x)\left[(m-1)x + (x - 4(m-1))\right]^2 - \left[(m-1)x - (x - 4(m-1))\right]^2 > 0$$
$$\iff 4(m-1)x(x - 4(m-1)) + x(mx - 4(m-1))^2 > 0$$
$$\iff 4(m-1)x - 16(m-1)^2 + m^2x^2 - 8mx(m-1) + 16(m-1)^2 > 0$$
$$\iff m^2x^2 + 4x(m-1)(1-2m) > 0$$
$$\iff m^2x^2 - 4x^2g(m-1) > 0$$
$$\iff m^2 > 4g(m-1).$$

# H  Numerical outcomes details: Fig. 6 and Fig. 7.

**Numerical setting.** The lower panel of Fig. 6 has been produced by running 3600 simulations, one for each couple of migration rate $m \in [0.01, 3]$ and intensity of selection $g \in [0.01, 3]$, for $t \leq T_{max} \in \left[\frac{300}{\varepsilon^2}, \frac{600}{\varepsilon^2}\right]$, with a criteria to cut the simulation short at a time greater than $\frac{300}{\varepsilon^2}$ if the difference between two consecutive steps is small enough. The value of the other parameters are the same for each simulation: $r = 1, \theta = 1, \kappa = 1, \varepsilon = 0.05$, as well as the initial state:

$$\begin{cases} n_1^0(z) = 0.99 \times \dfrac{e^{-\frac{(z-0.2)^2}{2\varepsilon^2}}}{\sqrt{2\pi}\varepsilon}, \\ n_2^0(z) = \dfrac{e^{-\frac{(z-0.2)^2}{2\varepsilon^2}}}{\sqrt{2\pi}\varepsilon}. \end{cases}$$

The initial state is taken as monomorphic, as the aim of this figure is to be compared to the theoretical outcomes that are predicted within the scope of the slow-fast analysis as stated in Theorem 3.1 (so when the initial state is close enough from the slow manifold).

**Scoring.** Each simulation final state $(n_1^f, n_2^f)$ is attributed a score between 0 and 1 according to the following scheme:

1. if $\max\left(N_1^f, N_2^f\right) < 0.01$, then the score is 0 (for extinction) and is corresponding to the deep purple color. Else, the score is a positive number (lower than 1) according to what follows.

2. if the variance in trait of the meta-population is greater than $2\varepsilon^2$, the score is 1 (corresponding to the color yellow). This would be the case if the final state is dimorphic, but more generally, this is to highlight the simulations whose final state does not fall in the small segregational variance regime analysis prediction (which in particular predicts that the distribution of trait in the meta-population is monomorphic (see Section 3), with a variance of order $\varepsilon^2$ (see (12)).

3. if both conditions above are not met, then the score $S$ is given according to the following formula:
$$S = \frac{5}{6} - \frac{1}{3}\frac{\left|N_2^f - N_1^f\right|}{N_1^f + N_2^f}.$$

This formula discriminates between symmetrical equilibria (who are characterized by equal population sizes, see Proposition 4.1), which typically have a score of $\frac{5}{6}$ (corresponding to the color light green), and asymmetrical equilibria, which have a discrepancy in local population sizes and therefore have a typically much lower score (in the blue tones).

**Adjustments for Fig. 7.** The methodology is the same for the lower panel of Fig. 6 and both panels of Fig. 7, at the exception of the initial state, set as:

$$\begin{cases} n_1^0(z) = 0.9 \times \dfrac{e^{-\frac{(z+1)^2}{2\varepsilon^2}}}{\sqrt{2\pi}\varepsilon}, \\ n_2^0(z) = \dfrac{e^{-\frac{(z-1)^2}{2\varepsilon^2}}}{\sqrt{2\pi}\varepsilon}. \end{cases}$$

and of the time step for the lower panel of Fig. 7, which is refined to keep up with the smaller value of $\varepsilon^2$.